\documentclass[11pt]{article}
\usepackage{fullpage}

\pdfoutput=1 

\newif\ifNatComm
\NatCommtrue

\newif\ifarXiv
\arXivtrue

\newif\ifNoSI

\ifNatComm
  \ifarXiv
    \renewcommand{\emph}[1]{\textit{#1}}
  \else
    \renewcommand{\emph}[1]{`#1'}
  \fi
  
  \usepackage{fancyhdr}
\fi

\usepackage{graphicx,xcolor}
\definecolor{footnoteColor}{rgb}{0.6471, 0.1098, 0.1882} 
\definecolor{linkColor}{rgb}{0., 0., 1.} 
\definecolor{urlColor}{rgb}{0., 0., 1.} 

\ifNatComm
  \ifarXiv
       \usepackage[colorlinks=true,citecolor=linkColor,linkcolor=linkColor,urlcolor=urlColor,breaklinks=true]{hyperref}
  \else
    \usepackage[hidelinks=true,breaklinks=true]{hyperref}
   \fi
\else
  \usepackage[colorlinks=true,citecolor=linkColor,linkcolor=linkColor,urlcolor=urlColor,breaklinks=true]{hyperref}
\fi

\usepackage{amssymb}
\usepackage{amsmath}
\graphicspath{
		{./figures/}
		{./figures/SI/}
             }
\usepackage{paralist}

\usepackage[font={footnotesize}]{caption}

\usepackage{paralist}

\usepackage{comment}
\newcommand{\vect}[1]{\mathbf{#1}}

\newcommand{\expfluc}{\tilde{\xi}} 
\newcommand{\expasymp}{\xi}        
\newcommand{\trel}{\tau_{\textrm p}}

\newcommand{\ulcase}{\lowercase} 
\newcommand{\subfig}[1]{\textrm{\ulcase{#1}}}
\newcommand{\subcap}[1]{(\textbf{\ulcase{#1}})} 
\newcommand{\capheader}[1]{\textbf{#1.}}
\newcommand{\refMethods}{\hyperref[sec:methods]{Methods section}}

\newcommand{\SInfo}{Supplementary Information} 
\newcommand{\SItem}{Supplementary } 
\newcommand*{\SIautoref}[1]{\hyperref[{#1}]{\SItem \autoref*{#1}}} 
\newcommand{\refSNotePrediction}{\hyperref[ssec:exp_separation_prediction]{\SItem Note 1}}
\newcommand{\refSNoteSmoothness}{\hyperref[ssec:smoothness_evidence]{\SItem Note 2}}
\newcommand{\refSNoteExperiments}{\hyperref[ssec:exp_separation_experiments]{\SItem Note 3}}

\ifNoSI
  \renewcommand{\refSNotePrediction}{{\SItem Note 1}}
  \renewcommand{\refSNoteSmoothness}{{\SItem Note 2}}
  \renewcommand{\refSNoteExperiments}{{\SItem Note 3}}
\fi

\newlength{\figureWidth}
\newlength{\SFigWidth}
\newlength{\SFigHalfWidth}
\newlength{\SFigWideWidth}

\usepackage[binary-units=true]{siunitx} 
\DeclareSIUnit\pixel{px}

\usepackage{tikz}
\usepackage{pgfplots}
\usetikzlibrary{positioning}
\usetikzlibrary{calc} 

\newcommand{\tikzLabel}[2]{
  \begin{tikzpicture}[font=\sffamily]
        \node[anchor=south west,inner sep=0] (image) at (0,0){#2};
            \begin{scope}[x={(image.south east)},y={(image.north west)}]
              \draw (0.01,0.9) node {\ulcase{#1}};
            \end{scope}
        \end{tikzpicture}
    }

\makeatletter

\usepackage{soul}
\usepackage{tikz}
\usetikzlibrary{calc}
\usetikzlibrary{decorations.pathmorphing}
  
\makeatletter
  
\newcommand{\defhighlighter}[3][]{%
    \tikzset{every highlighter/.style={color=#2, fill opacity=#3, #1}}%
}
  
\defhighlighter{green}{.35}

\newcommand{\highlight@DoHighlight}{
    \fill [ decoration = {random steps, amplitude=1pt, segment length=15pt}
          , outer sep = -15pt, inner sep = 0pt, decorate
          , every highlighter, this highlighter ]
          ($(begin highlight)+(0,8pt)$) rectangle ($(end highlight)+(0,-3pt)$) ;
}
  
\newcommand{\highlight@BeginHighlight}{
    \coordinate (begin highlight) at (0,0) ;
}
  
\newcommand{\highlight@EndHighlight}{
  \coordinate (end highlight) at (0,0) ;
}

\newdimen\highlight@previous
\newdimen\highlight@current

\DeclareRobustCommand*\highlight[1][]{%
  \tikzset{this highlighter/.style={#1}}%
  \SOUL@setup
  \def\SOUL@preamble{%
    \begin{tikzpicture}[overlay, remember picture]
      \highlight@BeginHighlight
      \highlight@EndHighlight
    \end{tikzpicture}%
  }%
  \def\SOUL@postamble{%
    \begin{tikzpicture}[overlay, remember picture]
      \highlight@EndHighlight
      \highlight@DoHighlight
    \end{tikzpicture}%
  }%
  \def\SOUL@everyhyphen{%
    \discretionary{%
      \SOUL@setkern\SOUL@hyphkern
      \SOUL@sethyphenchar
      \tikz[overlay, remember picture] \highlight@EndHighlight ;%
    }{%
    }{%
      \SOUL@setkern\SOUL@charkern
    }%
  }%
  \def\SOUL@everyexhyphen##1{%
    \SOUL@setkern\SOUL@hyphkern
    \hbox{##1}%
    \discretionary{%
      \tikz[overlay, remember picture] \highlight@EndHighlight ;%
    }{%
    }{%
      \SOUL@setkern\SOUL@charkern
    }%
  }%
  \def\SOUL@everysyllable{%
    \begin{tikzpicture}[overlay, remember picture]
      \path let \p0 = (begin highlight), \p1 = (0,0) in \pgfextra
        \global\highlight@previous=\y0
        \global\highlight@current =\y1
      \endpgfextra (0,0) ;
      \ifdim\highlight@current < \highlight@previous
        \highlight@DoHighlight
        \highlight@BeginHighlight
      \fi
    \end{tikzpicture}%
    \the\SOUL@syllable
    \tikz[overlay, remember picture] \highlight@EndHighlight ;%
  }%
  \SOUL@
}

\ifarXiv
  \else
  \ifNatComm
    \renewcommand{\highlight}[2][yellow]{\ul{#2}}
  \fi
\fi

\begin{document}

\newcommand{\titleText}{On the role of initial velocities in pair dispersion\\
			in a microfluidic chaotic flow}


\newcommand{\authorsInfoText}{Eldad Afik$^{1,2,\star}$ and Victor Steinberg$^{1}$ \\
			     {$^1$\small Department of Physics of Complex Systems} \\ 
			     {\small Weizmann Institute of Science} \\ 
			     {\small Rehovot 76100, Israel} \\
			     {$^2$\small Division of Biology and Biological Engineering} \\ 
			     {\small California Institute of Technology} \\
			     {\small Pasadena, CA 91125, USA} 
		     }

\newcommand{\doi}{\href{http://dx.doi.org/10.1038/s41467-017-00389-8}{10.1038/s41467-017-00389-8}}

\title{\titleText}

\author{\authorsInfoText}

\date{}

\maketitle

\begin{abstract}%
  Chaotic flows drive mixing and efficient transport in fluids, as well as the
  associated beautiful complex patterns familiar to us from our every day life
  experience. Generating such flows at small scales where viscosity takes over
  is highly challenging from both the theoretical and engineering perspectives.
  This can be overcome by introducing a minuscule amount of long flexible
  polymers, resulting in a chaotic flow dubbed \emph{elastic turbulence}.
  At the basis of the theoretical frameworks for its study lie the assumptions
  of a spatially smooth and random in time velocity field. Previous measurements
  of elastic turbulence have been limited to two-dimensions.
  Using a novel three-dimensional particle tracking method we conduct a
  microfluidic experiment, allowing us to explore elastic turbulence from the
  perspective of particles moving with the flow. Our findings show that the
  smoothness assumption breaks already at scales smaller than a tenth of the
  system size. Moreover, we provide conclusive experimental evidence that
  \emph{ballistic} separation prevails in the dynamics of pairs of tracers over
  long times and distances, exhibiting a memory of the initial separation
  velocities. The ballistic dispersion is universal, yet it has been overlooked
  so far in the context of small scales chaotic flows.
\end{abstract}


\maketitle

To truly appreciate how come many find elastic turbulence astonishing, we first
have to realise that our intuition is based on scenarios where the flow is
dominated by inertia, quantified by high values of the Reynolds number.
When we stir sugar in a cup of coffee we typically drive the liquid in circles using the
tea-spoon, yet the flow quickly evolves into a three-dimensional chaotic one,
tremendously accelerating the homogeneous distribution of the sweetener
throughout the beverage. This mixing flow is a manifestation of the non-linearity
due to the inertia of the fluid taking over the viscous dissipation; 
the ratio of the two is estimated by the Reynolds number.

Now imagine a fly walking in honey or a bacterium swimming in water --- one
cannot expect any dramatic effects on the flow beyond a few `bug' distance units
away from it.
When the typical velocities and length scales are small, corresponding to very
low values of the Reynolds number, the flows of non-complex liquids  -- also
known as \emph{Newtonian} -- are dominated by dissipation. As a result they can be
generically characterised as smooth and predictable. So long as the driving
force and the boundary conditions are steady, so will be the flow. A special
class of geometries can induce three-dimensional flows, which despite being
steady in time, may lead to mixing \cite{Ottino1990,Stroock2002,Simonnet2005}; 
these \emph{chaotic mixers} rely on patterned boundaries \cite{Stroock2002} or 
the vessel geometry \cite{Simonnet2005} to continuously generate recurring diverging
streamlines, which due to the low Reynolds remain fixed in space and time.
Therefore, mixing in microfluidic devices is normally limited to diffusion.

Nevertheless, when even a minute amount of long flexible polymers, such as
DNA and protein filaments, are introduced, the flow may develop a series of
elastic instabilities which render it irregular and twisted. This flow
--- \emph{elastic turbulence} \cite{Groisman2000,Groisman2001,Larson2000} ---
which is chaotic in time, has been shown to drive efficient mixing in
microfluidic devices as it can take place at extremely low values of the
Reynolds number \cite{Burghelea2004};
in the case of our experiment, more than six orders of magnitude smaller than the
critical value for inertial turbulence in a pipe \cite{Reynolds1883}. 
It is exactly for this reason that even a fluid dynamics expert may be amazed
when presented with the visual contrast between the mixing due to elastic
turbulence and the expected separation between fluid layers in a laminar flow
when the polymers are absent, as presented in Refs.~\cite[Fig.\,1]{Groisman2001} and
\cite[Figs.\,21--22]{Groisman2004}; more background on elastic turbulence can be
found in the review paper Ref.\,\cite{Steinberg2009} and the references
therein.

Understanding transport phenomena at small scales is of importance and wide
interest mainly for two reasons:
\ifNatComm
  first, %
\else
  \begin{inparaenum}[(i)]
     \item %
\fi
    much of the dynamics relevant for biology and chemistry takes place at
    these scales \cite{Groisman2001,Salazar2009,Celani2010,Woodhouse2013};
\ifNatComm
  second, %
\fi
    microfluidic devices are playing an important role in research and
    industrial technologies
    \cite{Khandurina2000,Stroock2002,deMello2006,Zhang2006,Sackmann2014},
    often including complex fluids and flows 
    whose dynamics still lack a universal description.
\ifNatComm\else
  \end{inparaenum}
\fi

To achieve a fundamental understanding of mixing and transport phenomena, these
need to be related and derived from their underlying microscopic level
of description, at its simplest, the dispersion of pairs of particles 
\cite{Falkovich2001,Bourgoin2006,Salazar2009}.
Inspired by seminal works on turbulence beneath the dissipative scale, 
theoretical attempts to understand elastic turbulence rely on the assumptions 
that the velocity field is smooth in space \cite{Fouxon2003,Berti2008,Steinberg2009},
associating it with the class known as the \emph{Batchelor regime}
\cite{Shraiman2000,Falkovich2001}.
For the dynamics of passive point-like tracers this means that the relative velocity
between pairs is proportional to the distance separating them, with the upshot of
exponential separation on average, asymptotically in time \cite{Falkovich2001,Salazar2009};
in \refSNotePrediction \ we sketch how the asymptotic exponential pair separation
prediction comes about.

The experimental study of pair separation dynamics in elastic turbulence, taking place
inside a tiny tube, has been limited thus far as it poses technical challenges: 
\ifNatComm
     first, the positions of tracers are needed to be resolved over long times and
       distances, in particular when the tracers get nearby to each other, whereas
       the flow is chaotic and three-dimensional;
     secondly, the scales at which the dynamics takes place require the use of a
       microscope, where three-dimensional imaging is non-trivial;
     thirdly, the flow fluctuations in time dictate a high temporal resolution; 
     and finally, the statistical nature of the problem demands a large sample of
       trajectories, which in turn requires long acquisition times and reliable
       automation.
\fi

To overcome these, we have implemented a novel method, which has been tested and
presented in Ref.\,\cite{Afik2015}.
In a nutshell, the three-dimensional positions of the fluorescent particles are determined from a single
camera two-dimensional imaging, by measuring the diffraction rings generated by the out-of-focus
particle; this way the particle localisation problem turns into a ring
detection problem, which is addressed accurately and efficiently in
Ref.\,\cite{Afik2015}.
By means of this direct Lagrangian particle tracking technique, we have established an
experimental database \cite{Afik2017b} of about $10^7$ trajectories derived from passive tracers
in elastic turbulence, generated in a curvilinear microfluidic tube; %
for further details see the \refMethods.

In this letter we report the results of pair dispersion due to the chaotic flow.
Our data reveals that the memory of the initial relative velocity prevails the
average dynamics, leading to a quadratic growth in time of the relative pair separation 
--- the so-called \emph{Ballistic dispersion} ---  
and shows no signature of the asymptotic exponential growth.
In addition we found that the relative velocity deviates from linear dependence on the
separation distance already at about 8\% of the tube width, indicating that the
linear velocity assumption is violated for the most part of the motion, in
contrast to the conceptual framework broadly used for the study of elastic
turbulence.


\section*{Results}

Let us consider a pair of passive tracers separated by the vector $\vect{R}$;
one realisation of such a pair is shown in \autoref{fig:intro}\subfig{a,b}.
The construction of the ensembles for the analysis to follow is outlined in 
\autoref{fig:intro} as well as in the \refMethods.

\subsection*{Establishing a statistically stationary elastic turbulence}

As our intuition builds upon the common day-to-day experience with high Reynolds
($Re >1$) flows, which are typically mixing, the chaotic nature of the
trajectories presented in \autoref{fig:intro} may escape many readers.
However, at the absence of polymers, the flow at low Reynolds ($Re < 1$) is laminar and
regular, and tracers maintain their distance from the channel boundaries, exhibiting no
crossing of trajectories; \cite[Figs.\,21--22]{Groisman2004} present the
striking contrast between the laminar case of the pure solvent and the mixing
elastic turbulence in the presence of polymers, both at low Reynolds.

Spatial features of the mean flow in our system, elastic turbulence in 
curvilinear microfluidic, can be revealed by transforming to the Eulerian frame of
reference, as presented in 
\ifNoSI
  \SItem Figure 2%
\else
  \SIautoref{fig:Spatial_Eulerian_mid_depth_slice}%
\fi%
, and
highlighted in its caption.
These are in accordance with two-dimensional Eulerian studies of
statistically stationary fully-developed elastic turbulence
\cite{Groisman2001,Jun2011}.
Despite some differences in the details of the experiments, this accordance
should come as no surprise since the numbers characterising our flow, a Reynolds
number smaller than $ 10^{-4}$ and a global Weissenberg number larger than 250
(see the \refMethods), indeed indicate that the results presented here were obtained
in a regime lying well beyond the critical values for its statistical scaling
properties to be Weissenberg and Reynolds dependent
\cite{Groisman2001,Jun2011}; that is, in our experiment the Reynolds number is small
enough to exclude any non linear effects due to inertia, and the Weissenberg number
is large enough to achieve the elastic turbulence flow state which is both
random in time and statistically time-independent.

A comparison of the local fluctuation intensities over time, as measured by the standard
deviation fields, to the magnitude of mean velocity components, supports the notion
of temporal randomness of our flow: this is most evident in the case of the
non-stream-wise velocity components which fluctuate over time to a degree which
exceeds that of the mean value in several regions across the pipe
cross-sections, and comparable even to the stream-wise velocity component;
see 
\ifNoSI
  \SItem Figure 2%
\else
  \SIautoref{fig:Spatial_Eulerian_mid_depth_slice}%
\fi%
, specifically
compare the values in sub-figures \subfig{e} to \subfig{c} and \subfig{f} to
\subfig{d}. Realisations of velocity fluctuations in time, highlighting the
randomness of the velocity field even at lower values of the Weissenberg number
($Wi$), have been shown in previous reports; 
see \cite[Fig.\,2]{Groisman2001} and \cite[Figs.\,16--17]{Jun2011} 
(when comparing, note that our flow parameters
should lead to a similar $Wi$  to the one in Ref.\,\cite{Groisman2001}, and are
close to the $Wi=679$ in Ref.\,\cite{Jun2011}; see the \refMethods\ for
clarification).

\ifarXiv
\begin{figure*}[htpb]
  \centering
  \begin{tikzpicture}[font=\sffamily]
        \node[anchor=south west,inner sep=0] (image) at (0,0)%
             {\includegraphics[width=.99\textwidth,trim=0 0 -1mm 27mm,clip]{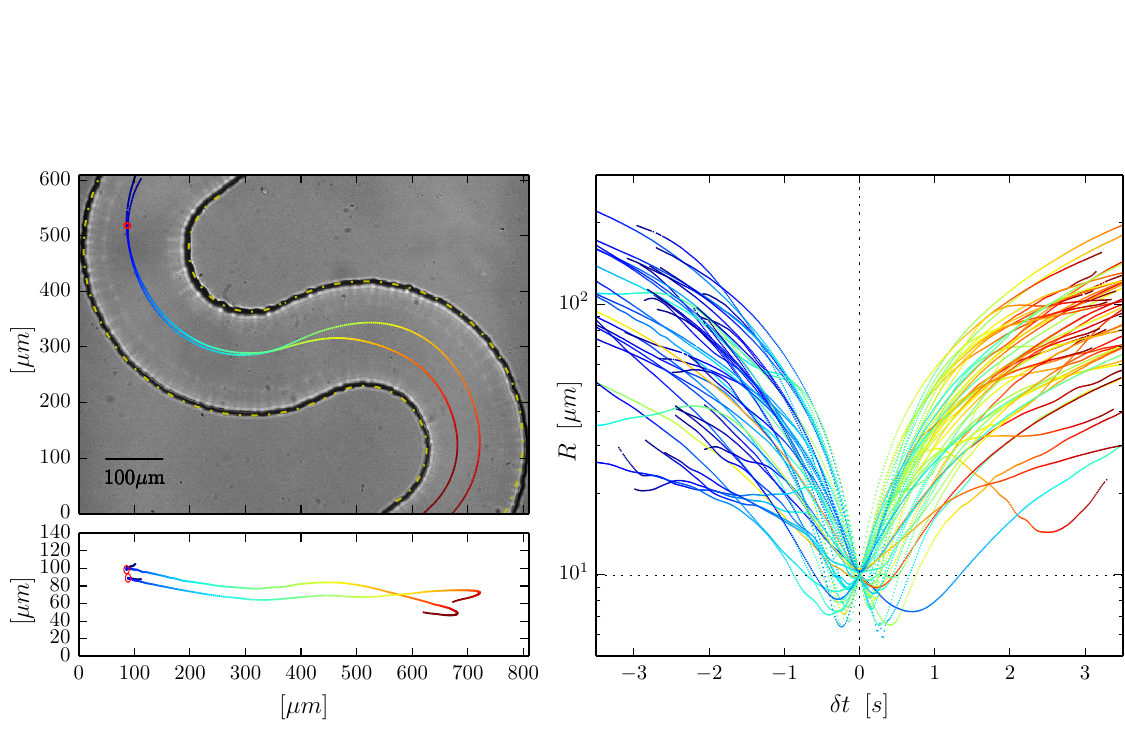}};
            \begin{scope}[x={(image.south east)},y={(image.north west)}]
              \draw (0.01,0.97) node {\ulcase{a}};
              \draw (0.01,0.37) node {\ulcase{b}};
              \draw (0.50,0.97) node {\ulcase{c}};
            \end{scope}
  \end{tikzpicture}
  \caption{\capheader{Pair dynamics example}
    The trajectories of two tracers are plotted in the left panels. The right
    panel shows a sub-sample of pair separation distances in the course of time. 
    The figure outlines the analysis forming the ensembles of pairs, as well as
    demonstrates the chaotic nature of the flow as manifested by pairs;
    to develop the intuition and contrast with laminar flow, the reader is
    referred to, e.g., \cite[Figs.\,21--22]{Groisman2004};
    several features of the mean flow in our case are manifested in
    the Eulerian representation in
    \ifNoSI
      \SItem Figure 2, %
    \else
      \SIautoref{fig:Spatial_Eulerian_mid_depth_slice}, %
    \fi%
    particularly the striking
    differences from Poiseuille-like laminar flows.
    \newline
    \subcap{a} A projection on to the plane of the camera, which is imaging the
    channel from the bottom side (gravity pointing out of the panel towards the
    reader), overlaid on a bright field image of the
    observation window (further technical details are provided in
    \ifNoSI
      \SItem Figure~1 %
    \else
      \SIautoref{fig:tube_geometry} %
    \fi%
     and in the \refMethods).
    \subcap{b} A side projection; the vertical axis is aligned with that of
    gravity as well as the channel depth, \SI{0}{\um} marking the channel bottom
    plane; as the width of this panel spans a spatial range which is nearly six
    times longer than its height, for the sake of visualisation the vertical
    axis is stretched by 3/2; the colour code in the plot denotes time, which
    spans \SI{4}{\second} in this case.
    \newline
    All pairs of tracers which were detected at some instant at a prescribed
    separation distance, $R_0 = 10 \pm 0.5$\,\si{\um} in this particular
    example, are collected to form one ensemble. The event at which the pair
    separation was nearest to $R_0$, marked by the red circles in the plot, is
    recorded as $t_0$ for the specific pair for later analysis.
    Each $R_0$ bin is \SI{1}{\um} wide and centred at $6$ through \SI{50}{\um},
    with sample sizes ranging from nearly $10^4$ to over $10^6$ pairs,
    respectively; sample size data are presented in
    \ifNoSI
      \SItem Figure 6. %
    \else
      \SIautoref{fig:sample_size}. %
    \fi%
    \newline
    \subcap{c} A sub-sample of pair separation distances 
    $R(\delta t)$ for 49 pairs belonging to the $R_0=\SI{10}{\um}$ ensemble, presented
    on a semi-logarithmic scale; For each pair, $\delta t = t-t_0$ is
    the time elapsed since $t_0$. The colour code denotes time, scaled
    separately for each curve. 
  }
    \label{fig:intro}
\end{figure*}
\else
\fi

\subsection*{Pursuing the asymptotic exponential pair dispersion}

Above we have recalled that random linear flows have been shown theoretically to
result in an asymptotic exponential pair dispersion
$ \langle R^2(\delta t) \rangle = R_0^2 \, \exp \left[ 2\expasymp\, \delta t\right] $
(%
\ifNoSI
  \SItem Equation 3 %
\else
  \SIautoref{eq:exp_prediction}
\fi%
 in \refSNotePrediction), 
where the exponential rate $\expasymp$ is independent of the initial separation $R_0$;
see \refSNotePrediction \ and references therein \cite{Falkovich2001,Salazar2009}.
It is worth noting that $2\expasymp$, which can be identified with the second
order generalised Lyapunov exponent, is not trivially related to the ordinary
maximal Lyapunov exponent in the generic case; see \cite[\textsection
3.2.1]{Paladin1987}, \cite[\textsection 5.3]{Cencini2010}, \cite{Falkovich2001}
and others. 
The evaluation of the asymptotic exponential rate $\expasymp$ has drawn much
attention in the literature;
references to theoretical and numerical surveys can be found towards the end of
\refSNotePrediction, while \refSNoteExperiments \ reviews the literature which
follows from previous experimental studies.
Our experimental data for 
$ \langle R^2(\delta t) \rangle_{R_0} \big/ R_0^2 $
is presented in \autoref{fig:normed_pair_dispersion} 
on a semi-logarithmic scale; 
here, and in all that follows, $\langle \dots \rangle_{R_0}$ denotes ensemble averages
differing by their initial separations $R_0$.
As discussed in the figure caption, our data shows no supporting evidence for
the exponential growth of $ \langle R^2(\delta t) \rangle $.

\ifarXiv
\begin{figure*}
  \centering
  \includegraphics[width=\figureWidth]{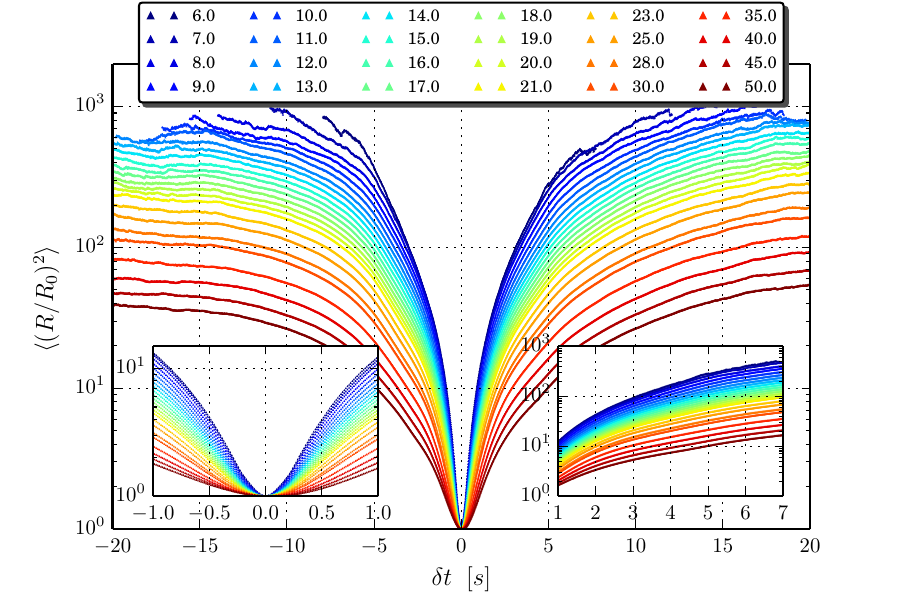}
  \caption{\capheader{Pair dispersion normalised by the initial separation}
    The plot shows the average squared pair separation distance, normalised by
    the initial separation, $\langle \left( R(\delta t) \big/ R_0 \right)^2
    \rangle_{R_0}$ for various $R_0$ between 6 and \SI{50}{\um}; 
    Curves satisfying the asymptotic exponential pair dispersion 
    $ \langle R^2(\delta t) \rangle = R_0^2 \, \exp \left[ 2\expasymp\,
    \delta t\right] $, 
    \ifNoSI
      \SItem Equation 3, 
    \else
      \SIautoref{eq:exp_prediction},
    \fi%
     would show-up on this
    semi-logarithmic presentation as straight lines, all sharing the same
    slope and, when extrapolated, hitting the origin, i.e., they should all
    collapse on a single linear relation.
    The insets present a zoom-in on the initial and intermediate temporal
    sub-intervals where the full range plot may seem to contain linear
    segments. Nevertheless, there is no unique slope which can be identified.
    Moreover, an exponential pair dispersion should extrapolate to the origin
    on this plot, which is clearly not the case here, and the curves do not
    merge asymptotically.
    The data shows no supporting evidence for the exponential time
    dependence which follows 
    \ifNoSI
      \SItem Equation 3.
    \else
      \SIautoref{eq:exp_prediction}.
    \fi%
    \newline
    The un-normalised data $\langle R^2 (\delta t) \rangle_{R_0}$ can be found in 
    \ifNoSI
      \SItem Figure 5. %
    \else
      \SIautoref{fig:not_normed_pair_dispersion}. %
    \fi%
  }
  \label{fig:normed_pair_dispersion}
\end{figure*}
\else
\fi

\subsection*{Failure of the linear flow assumption}

\ifarXiv
\begin{figure*}[htpb]
  \centering
  \includegraphics[width=\figureWidth]{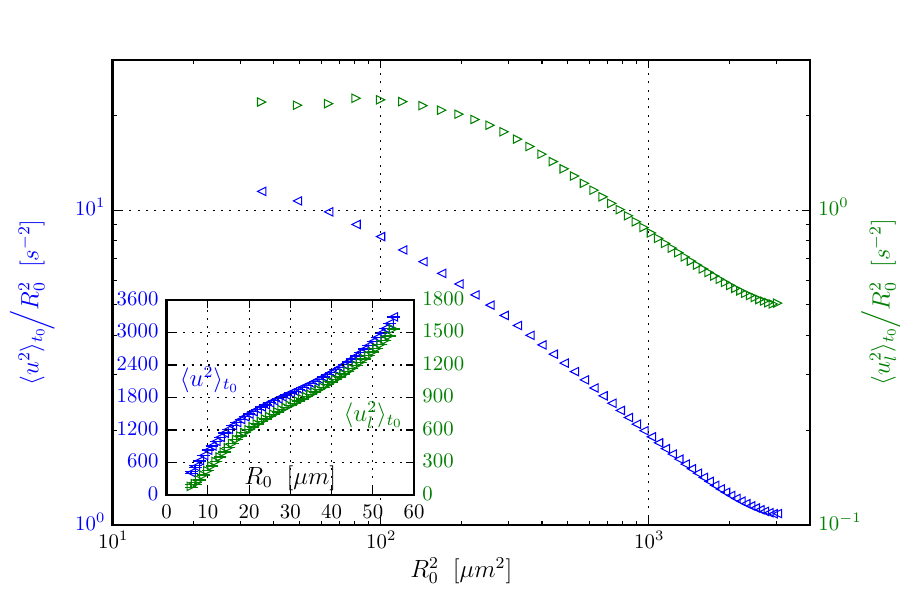}%
  \caption{\capheader{Initial relative velocity dependence on the separation
  distance} 
    The second moments of the relative velocity $\langle u^2 \rangle_{R_0,t_0}$ (blue
    left-triangles) and the separation velocity 
    $\langle u_l^2 \rangle_{R_0,t_0}$ (green right-triangle),  where 
    $u_l = \vect{u}\cdot\vect{R}\big/ R $,
    are plotted in the inset (right axis values are half the left ones) as
    function of the initial separation distance $R_0$; both ensemble averages
    are taken at the initial time $t_0$, when the pairs separation distance is
    closest to $R_0$.
    Rescaling these data by the squared initial separation $R_0^2$ 
    reveals the deviation from the commonly applied
    assumption of linear velocity field, as presented on a logarithmic scale in
    the main plot (right axis values are one order of magnitude
    smaller than the left ones).
    Had $\langle u^2 \rangle_R \propto R^2$ held, the rescaled curves would have
    remained constant; this is clearly not the case. 
    Indeed, the $\langle u_l^2\rangle_{R_0,t_0} / R_0^2$ data levels off
    as $R_0$ approaches the smaller distances, providing supporting evidence
    for the linearity of $u_l$ with $R$ at scales smaller than \SI{12}{\um}.
    However, this does not hold beyond a tenth of the channel depth.
    A linear flow regime is not supported by the rescaled relative
    velocity data $\langle u^2\rangle_{R_0,t_0} / R_0^2$, which values keep
    increasing even for the smallest $R_0$ values explored here.
        \newline
    Further note that $\langle u^2\rangle_{R_0,t_0}$ and 
    $\langle u_l^2\rangle_{R_0,t_0}$ (inset) are empirical estimators for the second order
    structure functions of the velocity and the longitudinal velocity,
    correspondingly; the former is the coefficient of the quadratic term in 
    \autoref{eq:Relative_dispersion_Taylor}.
    The error bars in the inset (smaller than the marker) indicate the margin
    of error based on a 95\% confidence interval.
  }
  \label{fig:u2_ul2_Ro_profile}
\end{figure*}
\else
\fi

This raises questions regarding the extent to which elastic turbulence can be
regarded as globally smooth, particularly in the presence of boundaries and
mean flow.
A velocity field consistent with linear flow behaviour would exhibit
$\langle u_l^2 \rangle_R \propto R^2$ for  the second order structure function
of the longitudinal velocity $\langle u_l^2\rangle_{R_0,t_0}$,
where $\vect{u}$ denotes the relative velocity and 
$u_l = \vect{u}\cdot\vect{R}\big/ R $; 
e.g. numerical simulations Ref.\,\cite[Figs.\,1 \& 6]{Frishman2015}.
In our flow, clear deviations from linearity are
evident already at separations beyond \SI{12}{\um}, less than 10\% of the
width and depth of the microfluidic channel, as can be learnt from
\autoref{fig:u2_ul2_Ro_profile}; a comparison to previous experimental results
is drawn in \refSNoteSmoothness.
The inset of \autoref{fig:u2_ul2_Ro_profile} presents the mean
squared relative velocities without rescaling; we shall return to these
profiles soon.

\subsection*{Relative pair dispersion}

Having not observed the exponential pair dispersion of long time asymptotics,
and noting that the pairs of tracers we study explore also regimes where the
linear flow assumption does not hold,
we were still left with the puzzle of the nature of the qualitative similarity among
the curves in \autoref{fig:normed_pair_dispersion} and its origin.
Using a different data-derived quantity we have found that, 
for a significant fraction of the observation time, the mean relative pair
dispersion evolves quadratically in time to leading order 
$\langle \|\vect{R}(\delta t)-\vect{R}_0 \|^2 \rangle \propto  \delta t ^2 $; 
this observation is evident in the insets of
\autoref{fig:short_time_behaviour_forward}. 
To better understand the source for this scaling let us
write the Taylor expansion around $\delta t = 0$
\begin{equation}
  \vect{R}(\delta t) = \vect{R}_0 + \vect{u}_0 \delta t +
  \frac{1}{2}\dot{\vect{u}}_0 \delta t^2 + \mathcal{O}\left(\delta t^3\right) \ .
  \label{eq:R_Taylor}
\end{equation}
Substituting this in the expression for the relative pair dispersion and
considering the ensemble average over pairs of the same initial separation
\begin{equation}
  \langle \| \vect{R}(\delta t) - \vect{R}_0 \|^2  \rangle_{R_0} =
  \langle u^2 \rangle_{R_0, t_0}\delta t^2 + 
  \langle \dot{\vect{u}}\cdot \vect{u}\rangle _{R_0, t_0} \delta t^3 +
  \mathcal{O}(\delta t^4) \ ,
  \label{eq:Relative_dispersion_Taylor}
\end{equation}
we find that the leading order term at short times is indeed quadratic in
$\delta t$ --- the so-called \emph{ballistic} regime.

\subsection*{Establishing the case for the short-time statistics}

To test this further we rescale the relative pair dispersion by the
pre-factor, the mean initial squared relative velocity 
$\langle u^2 \rangle_{R_0, t_0}$. Unlike the case of inertial turbulence,
for elastic turbulence there are no exact results nor scaling arguments to derive the
coefficients appearing in \autoref{eq:Relative_dispersion_Taylor}. 
Therefore we extract them from the experimental data;
see inset of \autoref{fig:u2_ul2_Ro_profile}.
Indeed, we find that our data admits a scaling collapse 
with no fitting parameters, providing a convincing experimental evidence that
these observations are well-described by the short time expansion of the
relative pair dispersion, exhibiting a significant deviation from $\delta t^2$ only after 2--3
seconds (see \autoref{fig:short_time_behaviour_forward}).

\ifarXiv
\begin{figure*}
  \centering
   \tikzLabel{a}{%
    \includegraphics[width=\figureWidth]{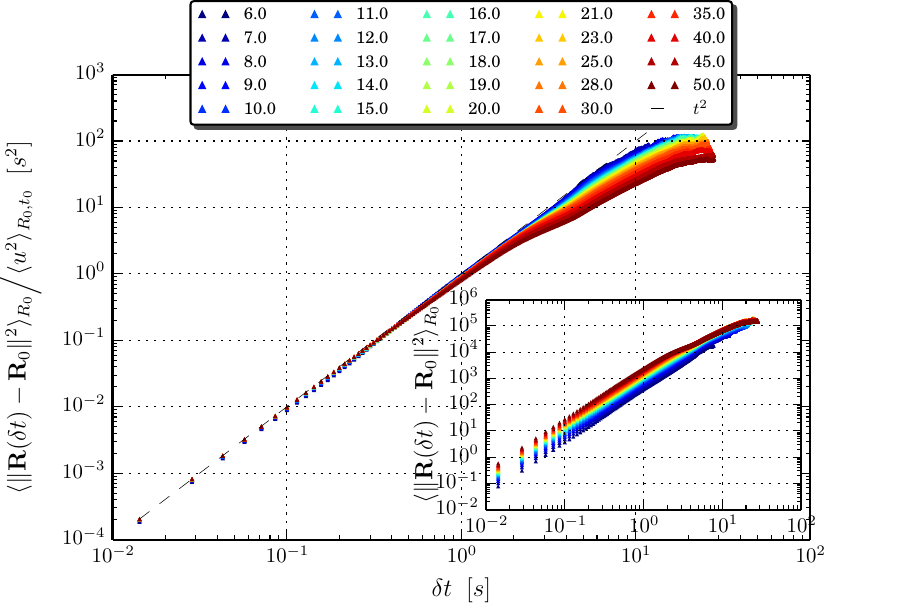}%
                }
   \begin{tikzpicture}[font=\sffamily]
        \node[anchor=south west,inner sep=0] (image) at (0,0)%
    {\includegraphics[width=\figureWidth,trim=0cm 0cm 0cm 0.66cm,clip]%
                             {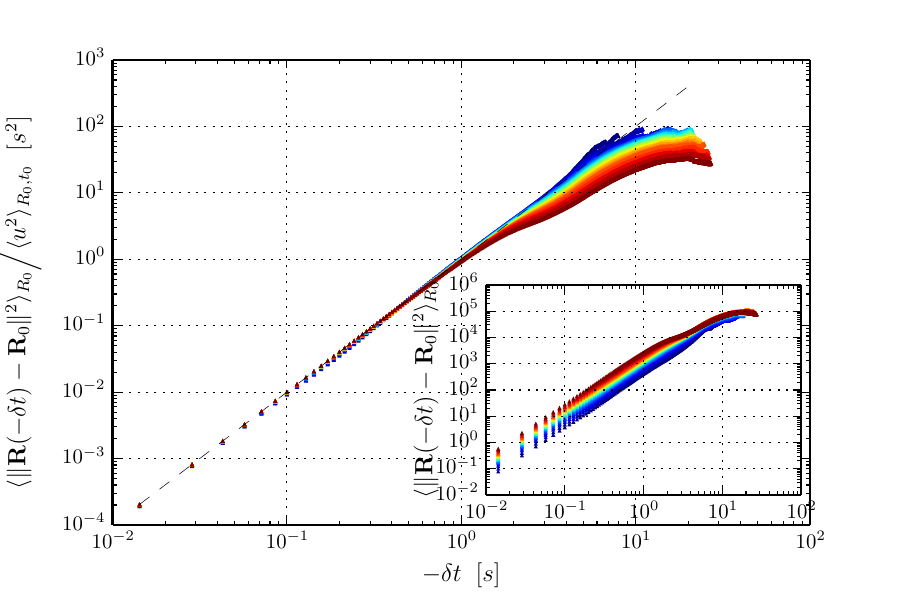}};
            \begin{scope}[x={(image.south east)},y={(image.north west)}]
              \draw (0.01,0.96) node {\ulcase{b}};
            \end{scope}
   \end{tikzpicture}
  \caption{\capheader{Relative pair dispersion forward and backwards in time
  evolutions}
    \newline
    \subcap{a} Forward in time $\langle \| \vect{R}(\delta t) - \vect{R}_0 \|^2
    \rangle_{R_0}$ for    various initial separations (inset) between 6 and
    \SI{50}{\um}, collapse initially on a single curve which follows a power-law
    $\delta t^2$, once rescaled by the average squared relative velocity at the
    initial time, $\langle u^2\rangle_{R_0, t_0}$.
    A significant deviation from $\delta t^2$ is noticed after
    2--\SI{3}{\second}, indicating 
    the time beyond which higher order terms should be considered.
    \subcap{b} Backwards in time relative pair dispersion 
    $\langle \| \vect{R}(- \delta t) - \vect{R}_0 \|^2  \rangle_{R_0}$ for the
    same initial separations (inset), showing the same initial scaling collapse as the
    forward in time. }
  \label{fig:short_time_behaviour_forward}
\end{figure*}
\else
\fi

Before discussing this time scale, we would like to first expose the
sub-leading contributions to the initial relative pair dispersion. To this end,
we subtract the backwards-in-time dynamics from the forward one.
This way the time-symmetric terms, even powers of $\delta t$, are eliminated.
The result, the time asymmetric contributions presented in
\autoref{fig:short_time_behaviour_backwards-forward}\subfig{a},
shows that indeed initially the next-to-leading order correction follows
$\delta t^3$ and that the curves do collapse onto a single one when rescaled by 
$\langle \dot{\vect{u}}\cdot \vect{u}\rangle$, the appropriate coefficient in
\autoref{eq:Relative_dispersion_Taylor}. 
The values of $\langle \dot{\vect{u}}\cdot \vect{u}\rangle_{R_0,t_0}$ were,
once again, extracted from the experimental data 
\ifNoSI
  \ (see \SItem Figure 3).
\else
  \ (see \SIautoref{fig:a.u_Ro_profile}).
\fi%

\ifarXiv
\begin{figure*}
  \centering
   \tikzLabel{a}{%
     \includegraphics[width=\figureWidth]{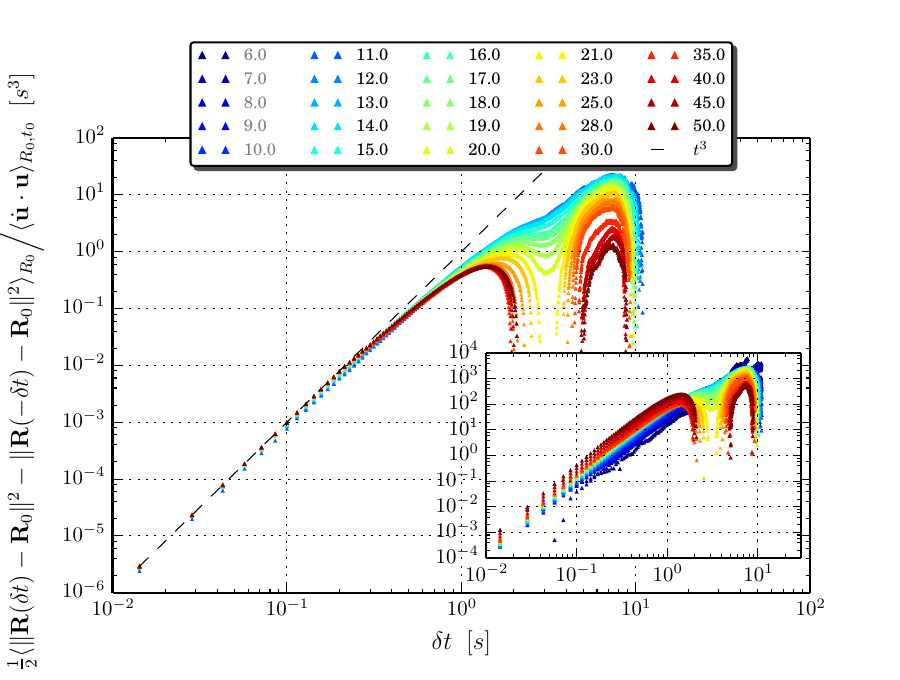}
   }
   \tikzLabel{b}{%
     \includegraphics[width=\figureWidth]{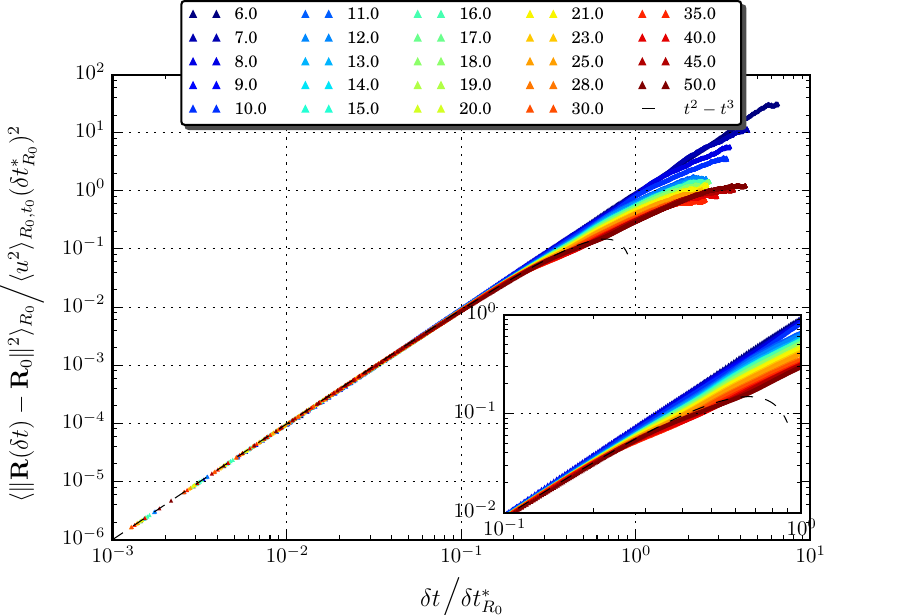}
   }
  \caption{\capheader{Relative pair dispersion time asymmetric terms and
  dimensionless form}
   \subcap{a}
    Taking the difference between the datasets plotted in the insets of
    \autoref{fig:short_time_behaviour_forward}, 
    $\frac{1}{2} \langle \| \vect{R}(\delta t ) - \vect{R}_0 \|^2 - 
                         \| \vect{R}(- \delta t ) - \vect{R}_0 \|^2
                         \rangle_{R_0}$,
    exposes the contribution of the time-asymmetric terms, odd powers in
    $\delta t$, presented here in the inset (sign inverted). 
    Rescaling by the empirical estimator for $\langle \dot{\vect{u}}\cdot \vect{u}
    \rangle_{R_0,t_0}$, these data collapse on $ \delta t^3 $ initially; 
    the datasets of $R_0 \leq \SI{10}{\um}$ (grey in the legend) are omitted from the
    main figure due to the scatter of the estimator; see 
    \ifNoSI
      \SItem Figure 3. %
    \else
      \SIautoref{fig:a.u_Ro_profile}. %
    \fi%
    The plot shows a deviation from $\delta t^3$ at times
    shorter than \SI{300}{\ms}, indicating the dominance of higher order (odd) terms at
    early times and that the $\delta t^3$ term alone does not trivially explain
    the deviation from $\delta t^2$, observed in
    \autoref{fig:short_time_behaviour_forward} after more than \SI{2}{\second}.
   \newline 
   \subcap{b}
    Rescaling the relative pair dispersion data (inset of
    \autoref{fig:short_time_behaviour_forward}\subfig{a}) by  
    $\langle u^2\rangle_{R_0, t_0} ( \delta t^*_{R_0} )^2$ 
    (see \autoref{eq:Relative_dispersion_Taylor}), 
    results in a dimensionless form, plotted here against dimensionless time,
    $\delta t$ rescaled by 
    $\delta t^*_{R_0} = \left| \langle u^2 \rangle_{R_0,t_0} \big/ \langle \dot{\vect{u}}\cdot
    \vect{u}\rangle_{R_0,t_0} \right|$; 
    the empirical estimators of $\delta t^*_{R_0}$ can be found in 
    \ifNoSI
      \SItem Figure 4. %
    \else
      \SIautoref{fig:dt_star_Ro_profile}. %
    \fi%
    The datasets indeed collapse onto a single curve
    $ (\delta t \big/ \delta t^*_{R_0})^2 - (\delta t \big/ \delta
    t^*_{R_0})^3$ (dashed black line) for $\delta t \big/ \delta
    t^*_{R_0} \lesssim 0.2$.
    The zoom-in (inset) emphasises the behaviour as $ \delta t \big/ \delta
    t^*_{R_0} $ approaches unity and the first two terms cancel out each other.
    }
  \label{fig:short_time_behaviour_backwards-forward}
  \label{fig:normalised_behaviour_forward}
\end{figure*}
\else
\fi

However, the deviations from this scaling are noticeable earlier than half a
second, much earlier than those from the ballistic behaviour
discussed above. This hints that the later deviation observed in
\autoref{fig:short_time_behaviour_forward} is in fact due to higher order
terms, potentially an indication of a transition to another regime.
The observation that this transition takes place at an earlier time for the larger
initial separations indicates the potential effects of the vessel size and its
geometry. It may also be attributed to the limited range of the linear flow approximation, consistent
with the data presented in \autoref{fig:u2_ul2_Ro_profile}.

\subsection*{Exploring how far the short-time statistics apply}

Finally, let us consider the limitations of the relative pair
dispersion short-time statistics description and its temporal range of
application.
The ratio of the first two coefficients in
\autoref{eq:Relative_dispersion_Taylor} constitute a time scale, 
$\delta t^*_{R_0} = \left| \langle u^2 \rangle_{R_0,t_0} \big/ \langle \dot{\vect{u}}\cdot
    \vect{u}\rangle_{R_0,t_0} \right|$%
, which puts an upper bound for the ballistic approximation to be relevant.
Rescaling \autoref{eq:Relative_dispersion_Taylor} by 
$\langle u^2\rangle_{R_0, t_0} ( \delta t^*_{R_0} )^2$, the equation attains a
dimensionless form, and one finds that the first two terms cancel each other as
$\delta t \big/ \delta t^*$ approaches unity due to the negative sign of 
$ \langle \dot{\vect{u}}\cdot \vect{u}\rangle_{R_0,t_0} $, 
giving place to higher order
terms to prevail the dynamics. 
Moreover, at that point, the expansion about the initial time is expected to
fail altogether.
The corresponding rescaled empirical data is presented in
\autoref{fig:normalised_behaviour_forward}\subfig{b}; 
the empirical $R_0$ profile of $\delta t^*_{R_0}$ is provided in
\ifNoSI
  \SItem Figure 4.
\else
  \SIautoref{fig:dt_star_Ro_profile}.
\fi%


\section*{Discussion}

On the one hand, our observations are consistent with the time-scale $\delta t^*$,
as sub-ballistic deviations from the $\delta t^2$ scaling are noticeable about
$\delta t \approx 0.1 \delta t^*$, as expected; 
see the zoom-in provided as the inset of \autoref{fig:normalised_behaviour_forward}, 
particularly for  $R_0 \gtrsim$\SI{23}{\um}, and to be compared with numerical
simulations of inertial turbulence \cite[Fig.\,1]{Bitane2012}.
On the other hand, the data indicates that the relative
pair dispersion remains near the $\delta t^2$ scaling even
when $\delta t \approx \delta t^*$, which is remarkable and puzzling.

A question that may naturally come to mind is whether one could match the two
limits, the short time-statistics and the long-time exponential prediction.
Before making any further observations, one has to recall that the two are
fundamentally different as the former is achieved by expanding about the intial
time while the latter is attained as time approaches infinity, so attempting to
match the two does not apply. 
Moreover, to demonstrate an exponential pair dispersion of the form of 
\ifNoSI
  \SItem Equation 3,
\else
  \SIautoref{eq:exp_prediction},
\fi%
 it is necessary to show that the pair separation
distance normalised by the initial separation, $\langle \left( R(\delta t) \big/
R_0 \right)^2 \rangle_{R_0}$, follows $ \exp \left[ 2\expasymp\, \delta t \right]
$ which is $R_0$ independent. Going back to
\autoref{fig:normed_pair_dispersion}, had our data supported the asymptotic
exponential dispersion, the curves should have appeared as straight lines in
this presentation, all having the same slope and, when extrapolated, hit the
origin, collapsing all on a single linear relation. 
Our results clearly rule out the exponential dispersion regime in this case.

Before closing we must note that the short time-statistics, leading to
ballistic dispersion, is a universal property which does not require any
assumptions on the character of the flow.
Thus far experimental \cite{Ouellette2006a} and numerical
\cite{Yeung2004,Bitane2012} results have been limited to the inertial subrange of
high Reynolds number turbulence.
Beneath the dissipative scale a sign of this behaviour has been observed in
simulations of inertial turbulence \cite[Fig.\,5]{Yeung2004}.

And yet, to our knowledge the ballistic dispersion regime has not been discussed
experimentally in the context of small scales chaotic flows, nor has it been
confronted with the exponential pair dispersion prediction
\ifNoSI
  \SItem Equation 3.
\else
  \SIautoref{eq:exp_prediction}.
\fi%
On the contrary, reading recent publications on the subject, namely
Refs.\,\cite{Jullien2003,Salazar2009,Ni2013}, one may come to believe that the
exponential dispersion has already been observed experimentally, while a closer
examination reveals that this is not the case; further elaboration can be found
in the \refSNoteExperiments \ and the conclusions therein.

We have demonstrated the predictive power of the ballistic dispersion in
microfluidics elastic turbulence, 
a model system for a broader class of bounded chaotic flows at small scales.


\section*{Methods}\label{sec:methods}
\subsection*{Methods Summary} 
The work presented here relies on constructing a database of trajectories in an
elastic turbulence flow \cite{Afik2017b}. Elastic turbulence is essentially a low Reynolds
number and a high Weissenberg number phenomenon. The former means the 
inertial non-linearity of the flow is over-damped by the viscous dissipation. The latter
estimates how dominant is the non-linear coupling of the elastic stresses to
the spatial gradients of the velocity field compared with the dissipation of these
stresses via relaxation. This is the leading consideration in the design of the flow cell.

The Lagrangian trajectories are inferred from passive tracers seeded in the
fluid. In order to study the dynamics of pairs, the three-dimensional positions of the tracers
are needed to be resolved, even when tracers get nearby to each other.
The requirement of large sample statistics dictates the long duration of the
experiment, which lasts over days. The fluctuations due to the chaotic nature
of the flow set the temporal resolution at milliseconds.
This leads to a data generation rate of about \SI{180}{\giga\byte\per\hour}. 
Hence both the acquisition and the analysis processes are required to be steady
and fully automated.
The three-dimensional positions of the fluorescent particles are determined using two-dimensional single
camera imaging, by measuring the diffraction rings generated by the
out-of-focus particle.
This way the particle localisation problem turns into a ring detection
problem. To this end a new algorithm has been developed and tested
\cite{Afik2015}; the source is freely available online  
(\url{https://github.com/eldad-a/ridge-directed-ring-detector}). 

\subsection*{Microfluidic apparatus}
The experiments were conducted in a microfluidic device, implemented in
polydimethylsiloxane elastomer by soft lithography, 
consisting of a curvilinear tube having a rectangular
cross-section. The depth is measured to be \SI{135}{\um}, the width is
approximately \SI{185}{\um} (see 
\ifNoSI
  \SItem Figure~1).
\else
  \SIautoref{fig:tube_geometry}).
\fi %
The geometry consists of a concatenation of 33 co-centric pairs of half circles.

The working fluid consists of polyacrylamide (MW=\SI{1.8e7}{\dalton} at mass fraction
of 80 parts per million) in aqueous sugar syrup (1:2 sucrose to d-sorbitol ratio;
mass fraction of 78\%), seeded with fluorescent particles (1 micron
Fluoresbrite YG Carboxylate particles, PolySciences Inc.) at number density of
about 50 tracers in the observation volume.  

The flow is gravity driven.

\subsection*{Physical considerations for the flow and passive tracers}
The viscosity of the Newtonian solvent, without the polymers, is estimated to be
1100 times larger than water viscosity at \SI{22}{\celsius}. This leads to a
polymer longest relaxation time of $\trel \simeq \SI{100}{\second}$
\cite{Liu2009}, which is the longest time scale characterising the relaxation of
elastic stresses in the solution.
The ratio of the fluorescent particles mass density to that the working
solution is about 0.75; Yet, the high viscosity of the working fluid and the
small radius of the particles qualify them as passive tracers -- the effects of
buoyancy and inertia are essentially negligible as the terminal velocity is of
the order of a tenth of a nanometer per second, and the inertia relaxation time is shorter
than the tenth of a nanosecond. Additionally, for all practical
purposes we are allowed to neglect altogether contributions from Brownian
motion to the dynamics of the fluorescent particles on the time scales over which
they are observed --- their diffusion coefficient leads to a variance increase
of about a micron-squared in an hour.

Local velocity averaged over time in the Eulerian frame of reference
showed a maximum over space of about  
$ \max_{\vect{x}} \overline{v(\vect{x},t)} \simeq \SI{250}{\um\per\second}$,
for $v(\vect{x},t)$ denoting instantaneous local fluid velocity, here inferred
from single particle trajectories, and time-averaging denoted by the bar.
This results in a Reynolds number $Re \lesssim 10^{-4}$ and a global Weissenberg number
$Wi = \trel \frac{\max_{\vect{x}} \overline{v(\vect{x},t)}}{\mathrm{width}/2} \gtrsim 250$.
To interpret these values in the light of Ref.~\cite{Jun2011}, one has to first match the
manner by which $Wi$ is estimated. Plugging in the values provided in that
report, using the maximal stream-wise velocity in \cite[Fig.\,10]{Jun2011}, in
the definition we use above, one finds that the maximal $Wi$ used in
Ref.~\cite{Jun2011} would correspond to \num{447} in our case; using
\cite[Fig.\,4]{Jun2011}, we can infer that the onset of developed elastic
turbulence corresponds to $Wi \simeq 165$, placing the parameters of our
experiment in the regime of statistically stationary fully developed elastic
turbulence.   

\subsection*{Imaging system}
The imaging system consists of an inverted fluorescence microscope (IMT-2,
Olympus), mounted with a Plan-Apochromat 20$\times$/0.8NA objective (Carl
Zeiss) and a fluorescence filter cube; a Royal-Blue LED (Luxeonstar) served
for the fluorophore excitation.  A CCD (GX1920, Allied Vision Technologies) was
mounted via zoom and 0.1$\times$ c-mount adapters (Vario-Orthomate 543513 and
543431, Leitz), sampling at \SI{70}{\Hz}, \SI{968}{\pixel}~$\times$~\SI{728}{\pixel}, 
covering \SI{810}{\um}~$\times$~\SI{610}{\um} laterally and the full depth of the tube. 
The camera control was based on a modification of the Motmot Python
camera interface package \cite{Straw2009}, expanded with a home-made
plug-in, to allow real-time image analysis in the RAM \cite{Afik2015}, recording only the
time-lapse positions of the tracers to the hard drive.

\subsection*{Lagrangian particle tracking}
To construct trajectories, the particle localisation procedure, introduced in
Ref.~\cite{Afik2015}, has to be complemented by a linking algorithm.
Here we implemented a kinematic model, in which future positions are inferred from the
already linked past positions. We used the code accompanying Ref.~\cite{Kelley2011}
as a starting point. The algorithm was rewritten in Python 
(primarily using SciPy \url{http://www.scipy.org/} \cite{SciPy}),
generalised to n-dimensions, the kinematic model modified to account for accelerations
as well, a memory feature was added to account for the occasional loss of
tracers, and it was optimised for better performance.
The procedure accounts for the physical process of particles advected by a
smooth chaotic flow and for the uncertainties.
These arise from the chaotic in time nature of the flow (`physical noise') as
well as from localisation and past linking errors (`experimental noise'). 
Finally, natural smoothing cubic splines are applied to smooth-out the
experimental noise and estimate the velocities and accelerations
\cite{Wasserm2007,Ahnert2007}.
The smoothing parameter is chosen automatically, where Vapnik's measure
takes the role of the usual generalised cross-validation, adapted from the
Octave splines package  \cite{Krakauer2014}.
Links to the corresponding open-source Python code are provided below, under
\hyperref[ssec:code_availability]{Code availability}. 

\subsection*{Pairs analysis}
Within the trajectories database we have identified pairs of tracers which were found at
some instant at a separation distance close to a prescribed initial separation
$R_0 = 6, 7, \ldots, \SI{50}{\um}$, to within $\delta R_0 = \pm \SI{0.5}{\um}$.
The initial time $t_0$ for a trajectory was set by the instant at which the
separation distance was closest to $R_0$. 
This way, each pair separation trajectory $R(\delta t)$  can
contribute to an $R_0$ pairs ensemble at most once. See \autoref{fig:intro}.
The number of pairs considered in each $R_0$ ensemble is plotted in
\ifNoSI
  \SItem Figure~6 %
\else
  \SIautoref{fig:sample_size} %
\fi%
as function of $\delta t$.

Examining the ensemble averages of the relative separation velocity at the
initial time $\langle u_l\rangle_{R_0,t_0}$, 
we do not find an indication that our sampling
method introduces a bias for converging or diverging trajectories, at least for 
$R_0 \lesssim \SI{22}{\um}$. 

Our data supports the linear flow approximation assumption at small enough
scales, as indicated by the ensemble averages of the initial relative separation
velocity; see \autoref{fig:u2_ul2_Ro_profile} where 
$\langle u_l^2 \rangle_{R_0,t_0} \big/ R_0^2 $ (green right-triangles) levels-off at $R_0
\lesssim \SI{12}{\um}$.
The same regime is not reached for the relative velocity, yet the 
$\langle u^2 \rangle_{R_0,t_0} \big/ R_0^2 $ data in 
\autoref{fig:u2_ul2_Ro_profile} (blue left-triangles) does not rule out this
possibility for smaller scales.

\subsection*{Code availability}\label{ssec:code_availability}
All programming and computer aided analysis in this work relies on open-source
projects; all based on tools from the SciPy ecosystem \cite{Perez2011},
primarily using IPython \cite{Perez2007} as an interactive computational
environment, Pandas \cite{McKinney2010} for data structures, and Matplotlib
\cite{Hunter2007} for plotting.

Much of the source code developed in the course of this study is 
available as open-source at:\\
 \url{https://github.com/eldad-a/ridge-directed-ring-detector} \\
 \url{https://github.com/eldad-a/particle-tracking} \\
 \url{https://github.com/eldad-a/natural-cubic-smoothing-splines} 

\subsection*{Data availability}
The data sets generated and analysed during the current study are available in
the figshare repository,
doi:\href{http://dx.doi.org/10.6084/m9.figshare.5112991}{10.6084/m9.figshare.5112991}
\cite{Afik2017b}.

\newpage
\ifNatComm
  \renewcommand{\emph}[1]{\textit{#1}}

  \renewcommand{\emph}[1]{`#1'}

\fi
\section*{Acknowledgements}
We thank A. Frishman for the helpful and extensive discussions of the
theory and O. Hirschberg for useful discussions of the mathematical and
statistical analysis; 
EA had fruitful discussions with J. Bec, S. Musacchio, D. Vincenzi, EW Saw and R. Chetrite, kindly
organised by the latter;
both authors gained from thorough discussions with V. Lebedev, as well as the helpful
reading and comments of an earlier version of the manuscript by G. Boffetta, 
A. Celani, and M. Feldman.
This work is supported by the Lower Saxony Ministry of Science and
Culture Cooperation (Germany; grant \#VWZN2729) and the Israel Science Foundation
(ISF; grant \#882/15).

\section*{Author contributions}
V.S. proposed the study of pair dispersion in elastic turbulence.
E.A. designed the experiment, performed the measurements and analysed the results.
Both authors discussed the results, the relevant literature, and wrote the manuscript.

\section*{Competing financial interests}
The author declares no competing financial interests.

\section*{Corresponding author}
Correspondence to \href{mailto:eldad.afik@gmail.com}{Eldad Afik}.

\pagebreak


\ifNoSI 
  \end{document}
\fi


\pagebreak

\makeatletter 

\pagenumbering{roman}
\renewcommand{\thepage}{SI--\@roman\c@page}

\renewcommand{\figurename}{\SItem Figure}
\setcounter{figure}{0}

\setcounter{equation}{0}

\ifNatComm
  \section*{{\SInfo}}
  \date{}

  \setlength{\footskip}{70pt}
  \pagestyle{fancy}

  \maketitle

  \renewcommand{\abstractname}{\vspace{-\baselineskip}} 
\fi

\subsection*{\SItem Figures}

\begin{figure*}[h]
  \centering
  \includegraphics[width=\figureWidth]{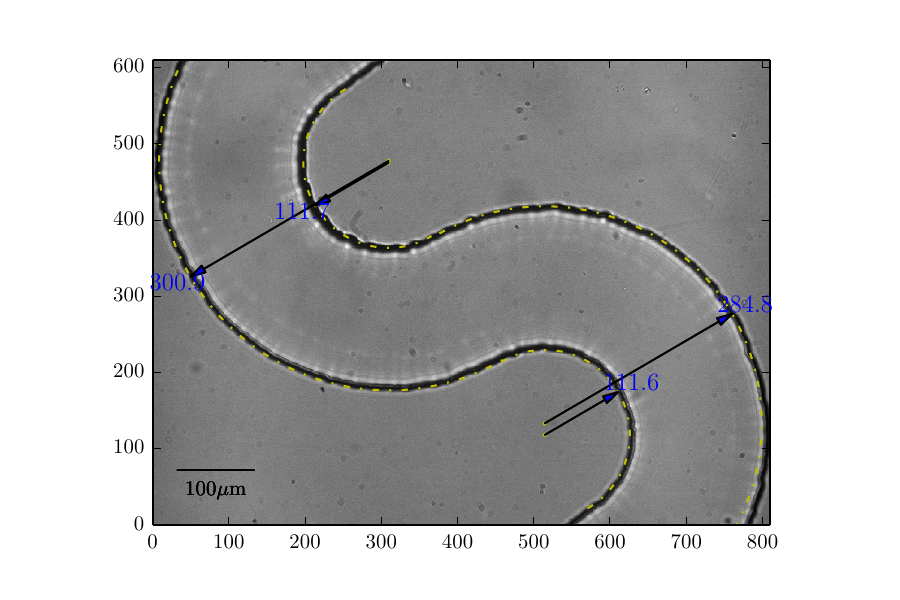}
  \caption{\capheader{Microfluidic tube geometry}
    A bright field image of the observation window; the tube geometry is based on a
    planar concatenation of co-centric pairs of half circles; the observation
    window encloses two out of 33 in total;  the arrows, and the yellow broken
    line indicate the tube walls and their dimensions, specified in microns; the
    tube has a rectangular cross-section of \SI{140}{\um} depth (perpendicular to
    the image plane). }
  \label{fig:tube_geometry}
\end{figure*}

\begin{figure*}
  \centering
   \tikzLabel{a}{%
	   \includegraphics[width=\SFigWideWidth]
	   					{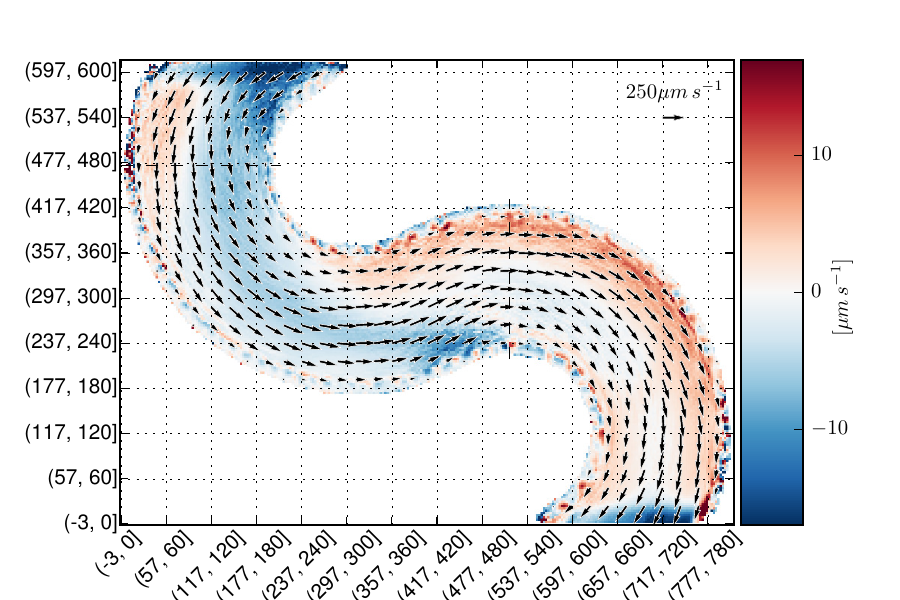}
   }

   \tikzLabel{b}{%
	   \includegraphics[width=\SFigWideWidth]
	   					{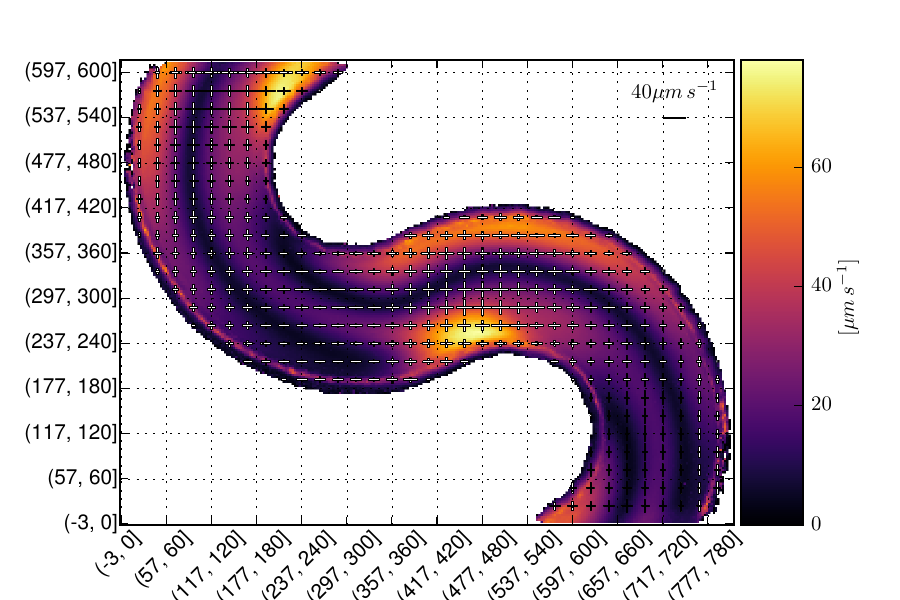}
   }
%
  \caption{\capheader{Spatial Eulerian statistics of the velocity field: mean
  value and standard-deviation over time}
  The transformation from the Lagrangian frame of reference to the Eulerian
  picture was achieved by subdividing the inner volume of the channel to a grid
  of \SI{3 x 3 x 3}{\um} units; data points along tracer trajectories were
  assigned to an Eulerian coordinate whenever they passed in the
  corresponding grid unit. The mean and standard-deviation of the velocity
  components for each unit sample were calculated. Here we
  present a selection of cross-sections along the channel, where the
  out-of-plane velocity component statistic is encoded in the colourmap; the
  standard-deviation of the in-plane components are represented by  line
  lengths.
  \subcap{a} \& \subcap{b} show the velocity field at mid-channel depth, mean and
  standard-deviation correspondingly; for the sake of visualisation, the in-plane 
  components are shown every eighth grid unit. The presence of an out of plane
  mean flow indicates an appreciable deviation from laminar Poiseuille-like
  flow; the cross-sections in what follows, corresponding to the two black dashed lines in
  \subcap{a}, further reveal the non-Poiseuille nature of the mean flow.
    }
  \label{fig:Spatial_Eulerian_mid_depth_slice}
\end{figure*}

\begin{figure*}\ContinuedFloat
  \centering
   \tikzLabel{c}{%
	   \includegraphics[width=\SFigHalfWidth]{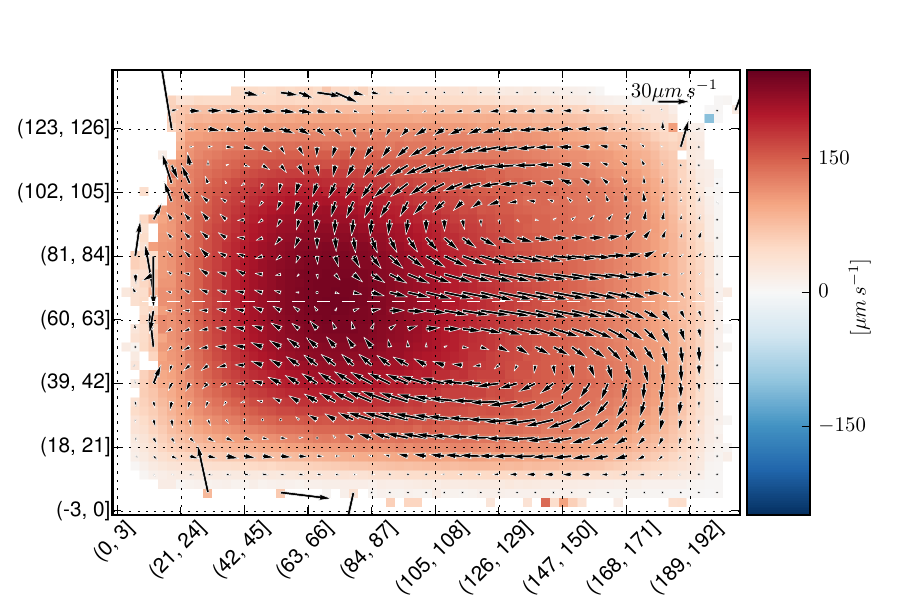}
   }%
   \tikzLabel{d}{%
	   \includegraphics[width=\SFigHalfWidth]{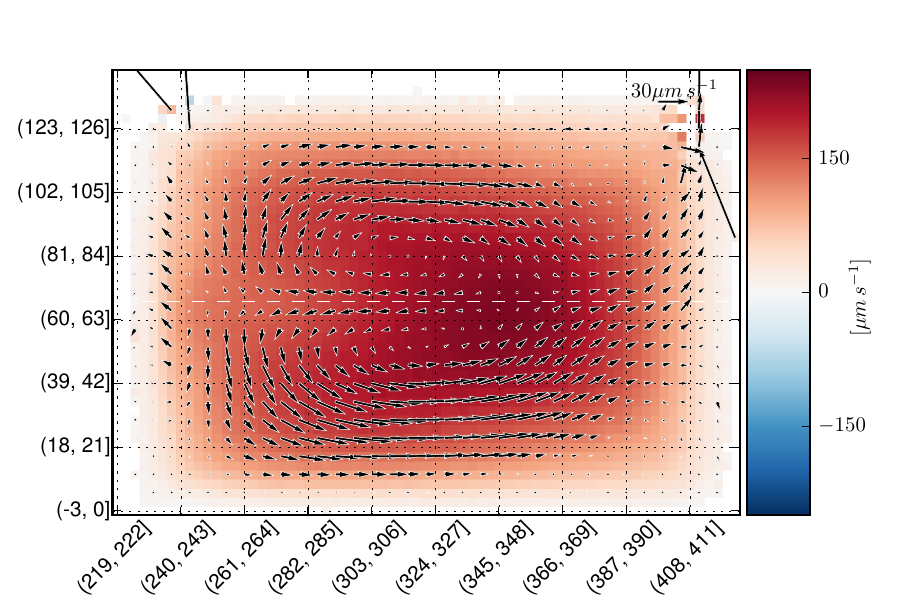}
   }

   \tikzLabel{e}{%
	   \includegraphics[width=\SFigHalfWidth]{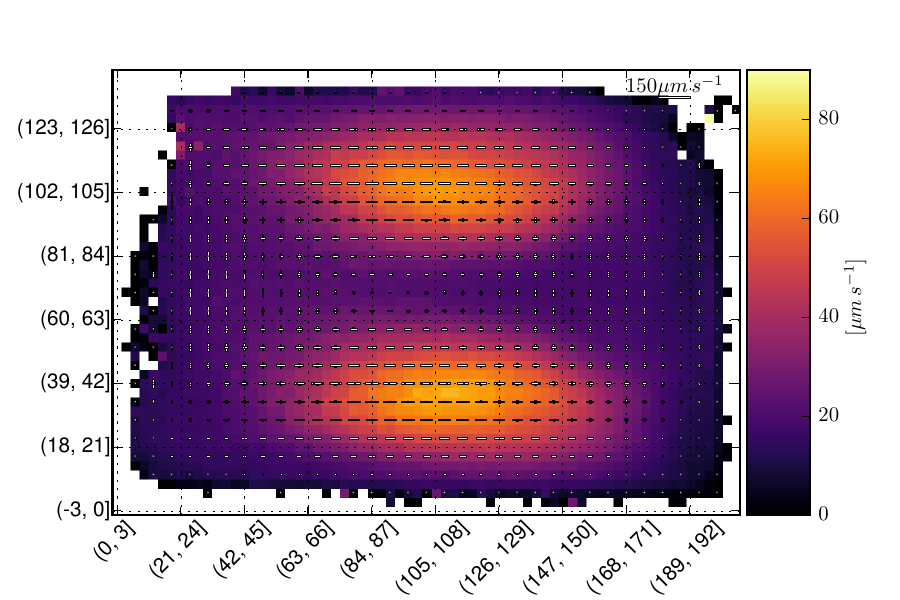}
   }%
   \tikzLabel{f}{%
	   \includegraphics[width=\SFigHalfWidth]{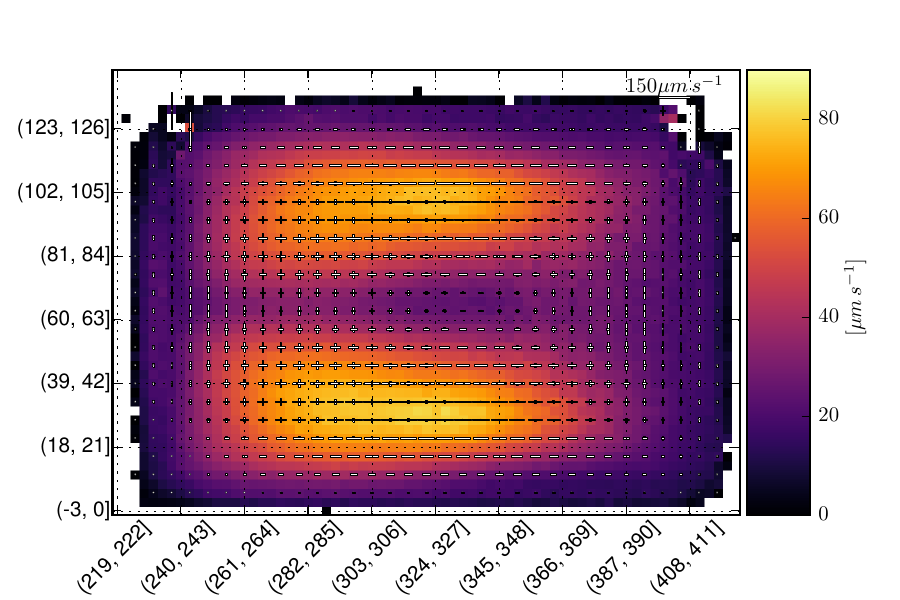}
   }
  \caption{\capheader{Spatial Eulerian statistics of the velocity field
	  (\textit{continued})}
	  \subcap{c} \& \subcap{e} present the velocity field on a
	  radial-vertical cross-section of the curvilinear channel, along the upstream dashed
	  line in \SIautoref{fig:Spatial_Eulerian_mid_depth_slice}\subfig{a}, mean
	  and standard-deviation correspondingly; \subcap{d} \& \subcap{f} show
	  the corresponding data for the downstream cross-section; the white
	  dashed lines in \subfig{c} \& \subfig{d} denotes the mid-channel depth, across which
	  \SIautoref{fig:Spatial_Eulerian_mid_depth_slice}\subfig{a} \& \subfig{b} is taken,
	  as well as the stream-wise component profiles to follow.
	  Both radial-vertical cross-section fields reveal
	  two main helical flows (circulating arrows; plotted every second grid
	  unit to ease visualisation) centred near the inner
	  wall, and a stream-wise component (colourmap) peak located
	  mid-way between them and shifted radially towards the external wall;
	  here inner and external refer to centres of the half circles
	  which form the tube, as shown in \SIautoref{fig:tube_geometry}.
	  The fluctuation intensities, as measured by the standard-deviation
	  fields, show higher values above and below the mid-depth line,
	  reaching values comparable to the mean velocity itself in extended
	  regions. The non-zero radial and vertical components themselves are in strong contrast to
	  what is expected had it been a laminar Poiseuille-like flow, where
	  these should vanish altogether.
    }
  \label{fig:Spatial_Eulerian_cross-sections}
\end{figure*}

\begin{figure*}\ContinuedFloat
  \centering
   \tikzLabel{g}{%
	   \includegraphics[width=\SFigHalfWidth]{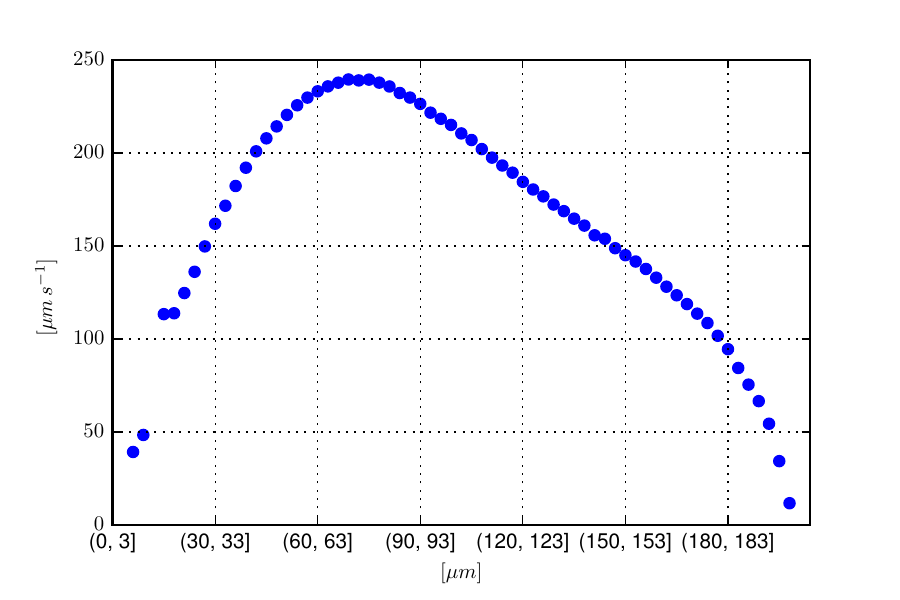}
   }%
   \tikzLabel{h}{%
	   \includegraphics[width=\SFigHalfWidth]{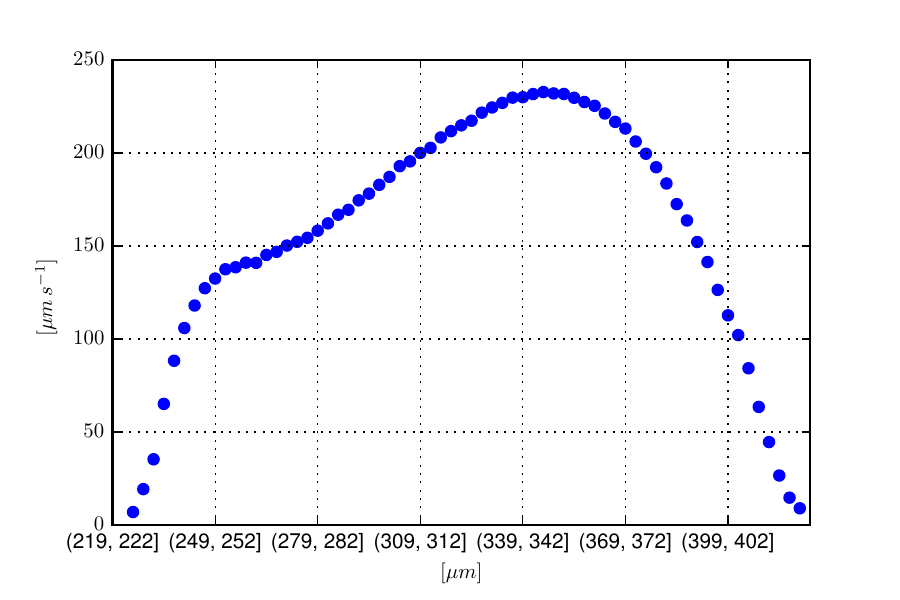}
   }
%
  \caption{\capheader{Spatial Eulerian statistics of the velocity field (\textit{continued})}
  \subcap{g} \& \subcap{h} show the stream-wise velocity component profile as function of
  the radial direction, along the mid-depth cuts, denoted by dashed white lines
  in \SIautoref{fig:Spatial_Eulerian_mid_depth_slice}\subfig{c} \& \subfig{d},
  correspondingly. As expected from developed elastic turbulence under similar
  geometry, these show high resemblence to the profile presented in
  \cite[Fig.\,10]{_Jun2011}; as can be seen, time-averaged longitudinal velocity
  exhibit a non-Poiseuille-like characteristic of an approximately linear
  profile over a significant fraction of the channel width, attributed to the
  efficient diffusion of momentum due to the mixing properties of elastic
  turbulence.
  These stand in sharp contrast with those found under the same conditions, only
  when polymers are absent, resulting in a laminar Poiseuille-like profile; see
  \cite[Fig.\,2]{_Jun2011}, where the laminar profile persisted even at values of
  the Reynolds number which were three orders of magnitude larger compared with
  the one in this study.
  }
  \label{fig:Spatial_Eulerian_cross-sections_profiles}
\end{figure*}

\begin{figure*}
  \centering
  \includegraphics[width=\SFigWidth]{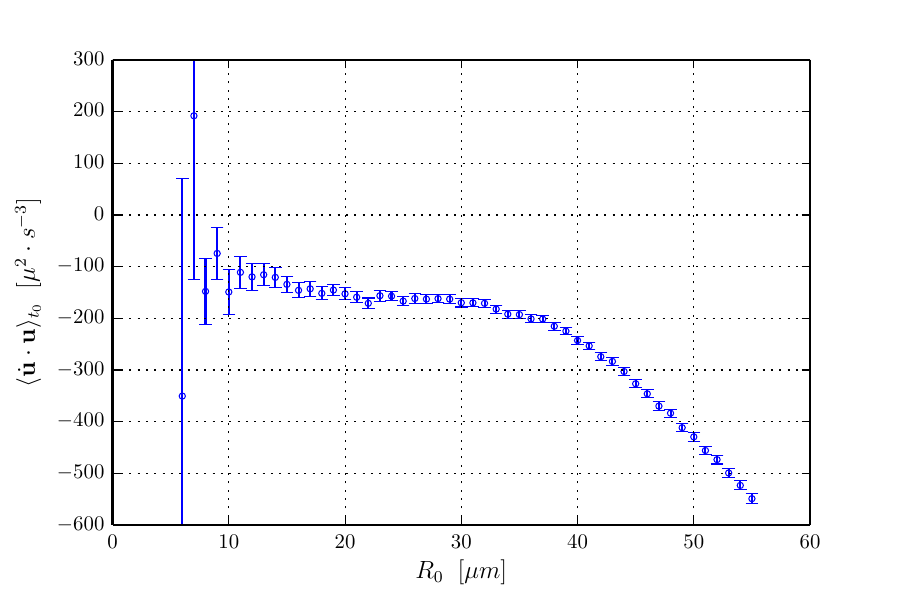}
  \caption{\capheader{The coefficient of the cubic term in
    \autoref{eq:Relative_dispersion_Taylor},
    $\langle \dot{\vect{u}}\cdot\vect{u}\rangle_{R_0,t_0}$}
    The averages are taken at $t_0$,  when the pairs separation distance is
    closest to $R_0$, as function of $R_0$, for  
    $\langle \dot{\vect{u}}\cdot\vect{u}\rangle_{R_0,t_0}$.
    It is also an estimator for the time derivative of $\langle \frac{1}{2}
    u^2\rangle_{R_0,t_0}$.
    The error bars indicate the margin of error based on the 95\% confidence. 
    }
  \label{fig:a.u_Ro_profile}
\end{figure*}

\begin{figure*}
  \centering
  \includegraphics[width=\SFigWidth]{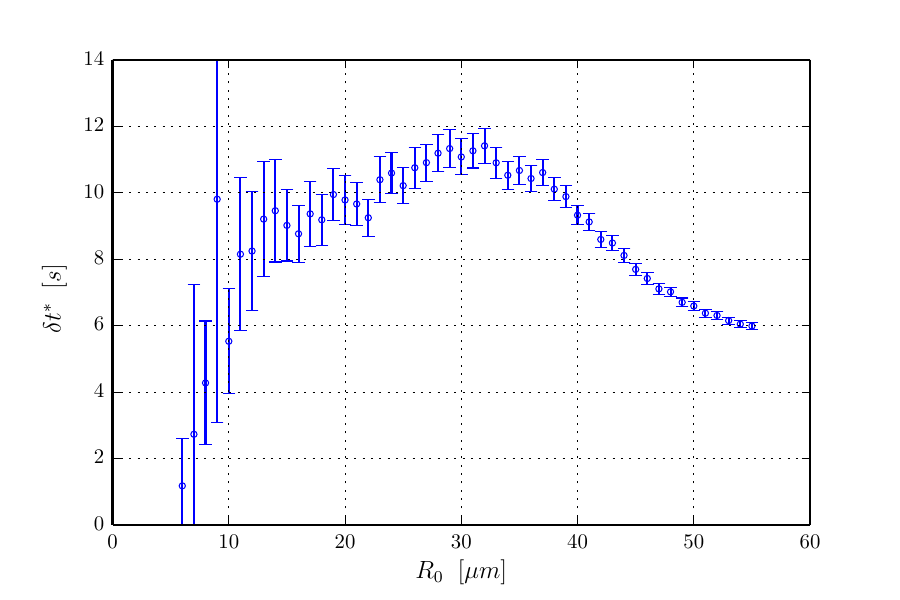}
  \caption{\capheader{Transition time from the ballistic regime}
    \newline
    $\delta t^*_{R_0} = \left| \langle u^2 \rangle_{R_0,t_0} \Big/ \langle \dot{\vect{u}}\cdot
    \vect{u}\rangle_{R_0,t_0} \right|$ is plotted as function of the initial separation
    distance $R_0$.
    Deviations of the relative dispersions 
    $\langle \| \vect{R}-\vect{R_0} \|^2 \rangle_{R_0}$ from the $\delta t^2$
    scaling are generically expected to be significant for $\delta t \gtrsim  0.1\, \delta t^*$.
    }
  \label{fig:dt_star_Ro_profile}
\end{figure*}

\begin{figure*}
  \centering
  \includegraphics[width=\SFigWidth]{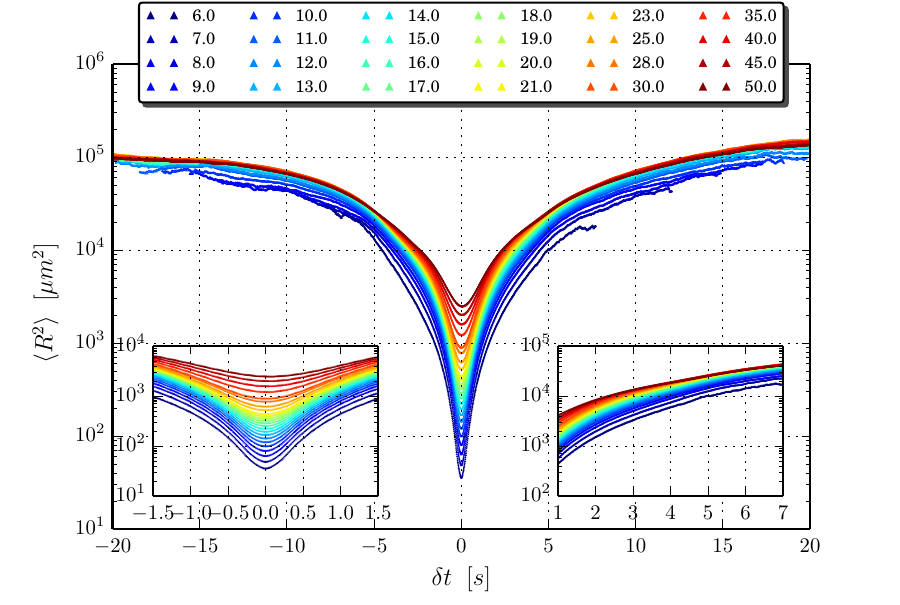}
  \caption{\capheader{Pair dispersion}
    The plot shows the average squared pair separation distance, 
    $\langle R^2\rangle_{R_0}$
    for various $R_0$ between 6 and \SI{50}{\um}; the insets show a zoom-in on
    the initial and intermediate time intervals.
    This is the same data plotted in the main text \autoref{fig:normed_pair_dispersion}%
    , only here not normalised by $R_0$.}
  \label{fig:not_normed_pair_dispersion}
\end{figure*}

\begin{figure*}
  \centering
  \includegraphics[width=\SFigWidth]{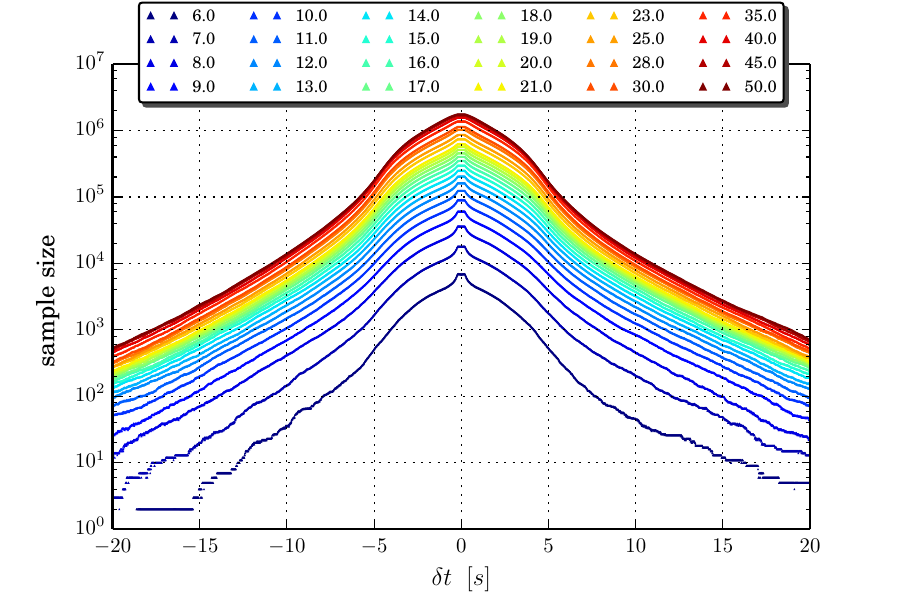}
  \caption{\capheader{Sample sizes}
    The number of pairs in the datasets presented in this work plotted
    against $\delta t$ for the various initial separation distances $R_0$.
    These sample sizes result from several filters on the observed pairs: each single
    trajectory is required to span more than \SI{140}{\um}; every $R(t)$
    trajectory is let to contribute to an $R_0$ bin at most once, at the point
    where it is nearest to the centre of the bin, this moment is denoted $t_0$
    for the pair in this bin; the pair trajectory is required to
    contain at least 15 measurement points before and after $t_0$. 
  Averages of sample sizes smaller than 100 were excluded.
  \label{fig:sample_size}}
\end{figure*}


\clearpage

\subsection*{\SItem Note 1: Exponential pair dispersion -- 
                            theoretical prediction}\label{ssec:exp_separation_prediction}

The study of particle dispersion in flows is at the basis of the understanding
of transport processes \cite{_Falkovich2001,_Bourgoin2006,_Salazar2009}.
In the early 1950s, G.~K. Batchelor predicted that the mean length of material lines in
turbulent flows would grow exponentially in the course of time, in the long
time limit; this stems from the notion that the line elements it consists of
can be considered short enough such that the distance between the ends of an
element remain within the dissipative scale throughout the motion
\cite{_Batchelor1952,_Monin2007b}.
In some recent works the terms \emph{material line} and \emph{material line
element} are used interchangeably to indicate the separation vector between two
passive particles in the fluid \cite{_Luthi2005,_Guala2005,_Salazar2009}.
And indeed Batchelor's prediction has been later reformulated for tracer particles in
the form of exponential pair separation; 
see \cite[\textsection 2]{_Salazar2009} for example.
To illustrate how this comes about, let us consider a pair of
passive tracers separated by the vector $\vect{R}$, 
and whose relative velocity is $\vect{u}$. The evolution of the
squared separation distance $R^2$ follows
\begin{equation}
  \frac{1}{2}\frac{d}{dt}R^2 = \vect{u}\cdot\vect{R} = u_l R \ ,
  \label{eq:R2_eom}
\end{equation}
where $u_l$ is the separation velocity, defined by the above relation.
The analysis then proceeds by assuming a linear flow approximation --- the
pair separation is considered to be small enough such that the velocity of one
tracer is linearly related to that of the other. Using this approximation 
\SIautoref{eq:R2_eom} reduces to 
\begin{equation}
  \frac{1}{2}\frac{d}{dt}R^2 =  \expfluc R^2 \ , 
  \label{eq:R2_linear}
\end{equation}
where $\expfluc$ no longer depends on $R$.
For the case of chaotic flows, $\expfluc$ can be modelled as a random variable, and
analyses often focus on the expectation value of such equations
\cite{_Monin2007b,_Falkovich2001,_Salazar2009}.
Additionally assuming the correlation time of $\expfluc$ to be very short compared to the
observation time \cite{_Falkovich2001,_Salazar2009}, a generalisation of the
central limit theorem -- the multiplicative ergodic theorem of Oseledec
\cite{_Falkovich2001,_Cencini2010} -- is applied, resulting in the exponential pair dispersion
\begin{equation}
  \langle R(t)^2\rangle = R^2(t_0)\, \exp \left\{ 2\expasymp\left(t-t_0 \right)
\right\} \  .
  \label{eq:exp_prediction}
\end{equation}
This is a relation for the time evolution of the second moment of pair separation
distances.
In this form one can identify $2\expasymp$ with the generalised Lyapunov exponent of the second 
order, which is generically not trivially related to the ordinary (maximal) Lyapunov 
exponent; see \cite[\textsection 3.2.1]{_Paladin1987}, \cite[\textsection
5.3]{_Cencini2010}, \cite{_Falkovich2001} and others.

Much of the theoretical and numerical literature discussing pair dispersion 
in the dissipative sub-range is devoted to the evaluation of $\expasymp$ in terms of
the typical time-scale of the flow $\tau$, that is $\expasymp = \gamma \big/
\tau$; see \cite[Eq.~2.9]{_Batchelor1952}. 
The value of $\gamma$ is still under debate as can be learnt from
\cite[\textsection 24.5]{_Monin2007b} as well as \cite{_Salazar2009} and references
therein.

\subsection*{\SItem Note 2: Smoothness of elastic turbulence
                            -- experimental evidence}\label{ssec:smoothness_evidence}

As we recall in the main text, the common theoretical framework employed to
analyse elastic turbulence relies on the assumption that the flow is smooth in
space \cite{_Fouxon2003,_Berti2008,_Steinberg2009}.
In this work we have demonstrated that the linear velocity field approximation does
not hold at scales beyond a rather small fraction of the
apparatus --- less than 10\% of the tube linear width in our case; see
\href{fig:u2_ul2_Ro_profile}{Figure 3}. This may seem at odds with previous
experimental reports \cite{_Burghelea2004a,_Burghelea2007,_Jun2011}.
Yet, we find no contradiction.

Based on the velocity power spectrum presented in
\cite[Fig.\,23]{_Burghelea2007},
Burghelea \textit{et al.} reported a power-law decaying faster than $-3$,
in support of the notion of a spatially smooth flow. 
However the lower cut-off for that scaling corresponds to scales
smaller than a third of the smallest spatial scale
of the apparatus, limiting the extent of the implications. 

The results presented in \cite[Fig.\,28--29 \& 33--34]{_Jun2011} are as
restrictive as ours.
There the scaling of structure functions changes before 10\% of the
apparatus size is reached.

One consequence of our findings is that the Lyapunov exponents picture
\cite{_Burghelea2004a,_Gerashchenko2005,_Berti2008} is not the appropriate
one to describe the dynamics of wall-bounded elastic turbulence at scales much
larger than few percent of the vessel size.
Indeed, Burghelea \textit{et al.} \cite{_Burghelea2004a} performed a Finite Time Lyapunov
Exponent analysis on numerically integrated particles (similarly to Jullien
\cite{_Jullien2003}), only one has to recognise that their results do not
demonstrate that in their experiment pairs diverged exponentially in time. 

This raises questions regarding the mechanism for polymer stretching once their
end-to-end distance goes beyond, as in our case.
In fact, Smith \textit{et al.} \cite{_Smith1999} have shown that a steady shear
flow is sufficient to stretch polymers to about 40\% of their full length.

\subsection*{\SItem Note 3: Exponential pair dispersion -- 
                            experimental evidence}\label{ssec:exp_separation_experiments}

\subsection*{Literature review}

At the time of writing, the exponential pair dispersion, briefly presented in
\refSNotePrediction, is regarded as the
leading paradigm for chaotic flows which are spatially smooth, as manifested by
the analysis of recent experimental results and the discussions which follow.

Jullien \cite{_Jullien2003} studied an instance of the Batchelor regime
flow in two-dimensional turbulence, where the velocity field was inferred experimentally
followed by numerical integration of tracers simulated on a computer;
the initial pair separation values were set to distances smaller than the
measurements grid.
An exponential separation, referred to in that context as Lin's law
\cite{_Lin1972}, was reported during an intermediate time interval of between
one to twice the value of the estimated flow typical time scale $\tau$, after which
a power-law scaling has been observed.

Salazar and Collins \cite{_Salazar2009} estimated $\gamma$ ($ = \expasymp \tau$) from measurements of
$\langle \expfluc \rangle$ in three-dimensional turbulence reported by Guala \textit{et al.}
\cite{_Guala2005}; 
this estimate should be taken with a grain of salt not only because it is unclear
whether these measurements were indeed restricted to the dissipative scales but
also as, although related, the quantities $\langle \expfluc \rangle$ and
$\expasymp$ are not the same \cite{_Monin2007b}.

Even more recently, Ni and Xia \cite{_Ni2013} reported measurements in 
three-dimensional turbulent thermal convection and inferred $\gamma$
from exponential fits to the mean squared pair separation distance; as
presented in \cite[Fig. 1]{_Ni2013}, the fits are taken at time intervals of up
to one Kolmogorov time-scale, a time too short with respect to the underlying
assumptions, and thereafter the data grows faster than the evaluated exponentials.
Additionally, as we mention in the main text, for an asymptotic exponential pair dispersion to be
demonstrated, the various curves should all have the same exponential rate and
differ only by the initial separation; such a result has not been empirically demonstrated.

We therefore find it safe to conclude that the literature on the subject is
lacking conclusive experimental evidence.

\subsection*{Discussion in the light of this study}
Studying pair separation in the dissipative sub-range over long times in
intense turbulence poses a technological challenge. The high velocities,
typical of high Reynolds number flows, restrict the length of the obtained
trajectories as exemplified by the above mentioned reports
\cite{_Jullien2003,_Luthi2005,_Ni2013} and other recent works \cite{_Bewley2013}.
This is one of the reasons for which the experimental literature on pair dispersion in
smooth chaotic flows is lagging behind the theoretical one.

Recall that the exponential growth in \SIautoref{eq:exp_prediction} relies on two underlying
assumptions \cite{_Batchelor1952,_Falkovich2001,_Salazar2009}: 
\begin{inparaenum}[(i)]
   \item the velocity field admits a linear approximation in space throughout
     the observation time; and
   \item the observation time is much longer than the correlation time of the
velocity gradients. 
\end{inparaenum}
These requirements are quite stringent and are clearly not fulfilled by our experiment.
To the best of our knowledge these assumptions have not been met experimentally
for tracer particles so far, and yet the exponential growth
prediction seems to be the leading paradigm in interpreting experimental
results \cite{_Jullien2003,_Salazar2009,_Ni2013}. 
Our estimations for the experimental system presented here indicate these may
be possibly relevant for 
$R_0 \lesssim \SI{0.1}{\um}$ and $\delta t \gtrsim \SI{10}{\second}$.
Examining the experimental parameters reported in recent works
\cite{_Bourgoin2006,_Ouellette2006a,_Bewley2013} we find that the trade-off between short
correlation times and a large enough dissipation scale renders them difficult
to reach in intense inertial turbulence.  

Following the above discussion one may reach the conclusion that the Batchelor prediction
\cite{_Batchelor1952} is irrelevant. By all means, this is not the case! 
Going back to Batchelor's own words: 
\begin{quote}
\textit{
      ``In that paper no consideration was given to the particular case of two particles
      which are so close together -- or of clouds whose linear dimensions are
      so small --
      that the value of the spatial derivative of the velocity is the same at the (simultaneous) 
      positions of the particles, for the reason that 
      \highlight[yellow]{such a condition is unlikely to be
      realized with practical methods of marking and observing particular fluid
      particles}.}

    \dots

\textit{
      Difficulties of observation of marked particles which are very close together do not
      worry us in the present connexion, since \highlight[yellow]{we are concerned here with the changes
      in the total length of a material line}, which is made up of the changes in a large
      number of infinitesimal line elements, unlike the changes in the shortest distance
      between two fluid particles.''
    }

  \hfill{\raggedleft{--- see G.~K. Batchelor, \textit{Proc. R. Soc. A} 1952
  \cite{_Batchelor1952}}}
\end{quote}
We find that Batchelor himself appreciated how challenging would his
assumptions be when it comes to finite size particles. 
At the same time, it is important to note that his prediction addresses the
evolution of the total length of a material line, where each sub-segment could
always be taken infinitesimally small, unlike the case of the shortest distance
between pairs of tracers starting with a finite initial separation.

\newpage
\ifNatComm
  \renewcommand{\emph}[1]{\textit{#1}}
  \renewcommand{\refname}{Supplementary References}

  \renewcommand{\emph}[1]{`#1'}
\fi


\begin{thebibliography}{10}
     \expandafter\ifx\csname url\endcsname\relax
       \def\url#1{\texttt{#1}}\fi
     \expandafter\ifx\csname urlprefix\endcsname\relax\def\urlprefix{URL }\fi
     \providecommand{\bibinfo}[2]{#2}
     \providecommand{\eprint}[2][]{\url{#2}}
     
     \bibitem{Ottino1990}
     \bibinfo{author}{Ottino, J.~M.}
     \newblock \bibinfo{title}{Mixing, chaotic advection, and turbulence}.
     \newblock \emph{\bibinfo{journal}{Annu. Rev. Fluid Mech. }}
       \textbf{\bibinfo{volume}{22}}, \bibinfo{pages}{207--254}
       (\bibinfo{year}{1990}).
     
     \bibitem{Stroock2002}
     \bibinfo{author}{Stroock, A.~D.} \emph{et~al.}
     \newblock \bibinfo{title}{Chaotic mixer for microchannels}.
     \newblock \emph{\bibinfo{journal}{Science}} \textbf{\bibinfo{volume}{295}},
       \bibinfo{pages}{647--651} (\bibinfo{year}{2002}).
     
     \bibitem{Simonnet2005}
     \bibinfo{author}{Simonnet, C.} \& \bibinfo{author}{Groisman, A.}
     \newblock \bibinfo{title}{Chaotic mixing in a steady flow in a microchannel}.
     \newblock \emph{\bibinfo{journal}{Phys. Rev. Lett. }}
       \textbf{\bibinfo{volume}{94}} (\bibinfo{year}{2005}).
     
     \bibitem{Groisman2000}
     \bibinfo{author}{Groisman, A.} \& \bibinfo{author}{Steinberg, V.}
     \newblock \bibinfo{title}{Elastic turbulence in a polymer solution flow}.
     \newblock \emph{\bibinfo{journal}{Nature}} \textbf{\bibinfo{volume}{405}},
       \bibinfo{pages}{53--55} (\bibinfo{year}{2000}).
     
     \bibitem{Groisman2001}
     \bibinfo{author}{Groisman, A.} \& \bibinfo{author}{Steinberg, V.}
     \newblock \bibinfo{title}{Efficient mixing at low reynolds numbers using
       polymer additives}.
     \newblock \emph{\bibinfo{journal}{Nature}} \textbf{\bibinfo{volume}{410}},
       \bibinfo{pages}{905--908} (\bibinfo{year}{2001}).
     
     \bibitem{Larson2000}
     \bibinfo{author}{Larson, R.~G.}
     \newblock \bibinfo{title}{Fluid dynamics: Turbulence without inertia}.
     \newblock \emph{\bibinfo{journal}{Nature}} \textbf{\bibinfo{volume}{405}},
       \bibinfo{pages}{27--28} (\bibinfo{year}{2000}).
     
     \bibitem{Burghelea2004}
     \bibinfo{author}{Burghelea, T.}, \bibinfo{author}{Segre, E.},
       \bibinfo{author}{Bar-Joseph, I.}, \bibinfo{author}{Groisman, A.} \&
       \bibinfo{author}{Steinberg, V.}
     \newblock \bibinfo{title}{Chaotic flow and efficient mixing in a microchannel
       with a polymer solution}.
     \newblock \emph{\bibinfo{journal}{Phys. Rev. E}} \textbf{\bibinfo{volume}{69}},
       \bibinfo{pages}{066305} (\bibinfo{year}{2004}).
     
     \bibitem{Reynolds1883}
     \bibinfo{author}{Reynolds, O.}
     \newblock \bibinfo{title}{An experimental investigation of the circumstances
       which determine whether the motion of water shall be direct or sinuous, and
       of the law of resistance in parallel channels.}
     \newblock \emph{\bibinfo{journal}{Phil. Trans. R. Soc. Lond. }}
       \textbf{\bibinfo{volume}{174}}, \bibinfo{pages}{935--982}
       (\bibinfo{year}{1883}).
     
     \bibitem{Groisman2004}
     \bibinfo{author}{Groisman, A.} \& \bibinfo{author}{Steinberg, V.}
     \newblock \bibinfo{title}{Elastic turbulence in curvilinear flows of polymer
       solutions}.
       \newblock \emph{\bibinfo{journal}{New J. Phys. }}
       \textbf{\bibinfo{volume}{6}}, \bibinfo{pages}{29} (\bibinfo{year}{2004}).
     
     \bibitem{Steinberg2009}
     \bibinfo{author}{Steinberg, V.}
     \newblock \bibinfo{title}{Elastic stresses in random flow of a dilute polymer
       solution and the turbulent drag reduction problem}.
       \newblock \emph{\bibinfo{journal}{C R Phys}}
       \textbf{\bibinfo{volume}{10}}, \bibinfo{pages}{728--739}
       (\bibinfo{year}{2009}).
     
     \bibitem{Salazar2009}
     \bibinfo{author}{Salazar, J.~P.} \& \bibinfo{author}{Collins, L.~R.}
     \newblock \bibinfo{title}{Two-particle dispersion in isotropic turbulent
       flows}.
     \newblock \emph{\bibinfo{journal}{Annu. Rev. Fluid Mech. }}
       \textbf{\bibinfo{volume}{41}}, \bibinfo{pages}{405--432}
       (\bibinfo{year}{2009}).
     
     \bibitem{Celani2010}
     \bibinfo{author}{Celani, A.} \& \bibinfo{author}{Vergassola, M.}
     \newblock \bibinfo{title}{Bacterial strategies for chemotaxis response}.
     \newblock \emph{\bibinfo{journal}{Proc. Natl. Acad. Sci. U.S.A. }}
       \textbf{\bibinfo{volume}{107}}, \bibinfo{pages}{1391--1396}
       (\bibinfo{year}{2010}).
     
     \bibitem{Woodhouse2013}
     \bibinfo{author}{Woodhouse, F.~G.} \& \bibinfo{author}{Goldstein, R.~E.}
     \newblock \bibinfo{title}{Cytoplasmic streaming in plant cells emerges
       naturally by microfilament self-organization}.
     \newblock \emph{\bibinfo{journal}{Proc. Natl. Acad. Sci. U.S.A. }}
       \textbf{\bibinfo{volume}{110}}, \bibinfo{pages}{14132--14137}
       (\bibinfo{year}{2013}).
     
     \bibitem{Khandurina2000}
     \bibinfo{author}{Khandurina, J.} \emph{et~al.}
     \newblock \bibinfo{title}{Integrated system for rapid {PCR}-based {DNA}
       analysis in microfluidic devices}.
       \newblock \emph{\bibinfo{journal}{Anal. Chem. }}
       \textbf{\bibinfo{volume}{72}}, \bibinfo{pages}{2995--3000}
       (\bibinfo{year}{2000}).
     
     \bibitem{deMello2006}
     \bibinfo{author}{deMello, A.~J.}
     \newblock \bibinfo{title}{Control and detection of chemical reactions in
       microfluidic systems}.
     \newblock \emph{\bibinfo{journal}{Nature}} \textbf{\bibinfo{volume}{442}},
       \bibinfo{pages}{394--402} (\bibinfo{year}{2006}).
     
     \bibitem{Zhang2006}
     \bibinfo{author}{Zhang, C.}, \bibinfo{author}{Xu, J.}, \bibinfo{author}{Ma, W.}
       \& \bibinfo{author}{Zheng, W.}
     \newblock \bibinfo{title}{{PCR} microfluidic devices for {DNA} amplification}.
     \newblock \emph{\bibinfo{journal}{Biotechnol. Adv. }}
       \textbf{\bibinfo{volume}{24}}, \bibinfo{pages}{243--284}
       (\bibinfo{year}{2006}).
     
     \bibitem{Sackmann2014}
     \bibinfo{author}{Sackmann, E.~K.}, \bibinfo{author}{Fulton, A.~L.} \&
       \bibinfo{author}{Beebe, D.~J.}
     \newblock \bibinfo{title}{The present and future role of microfluidics in
       biomedical research}.
     \newblock \emph{\bibinfo{journal}{Nature}} \textbf{\bibinfo{volume}{507}},
       \bibinfo{pages}{181--189} (\bibinfo{year}{2014}).
     
     \bibitem{Falkovich2001}
     \bibinfo{author}{Falkovich, G.}, \bibinfo{author}{Gaw\c{e}dzki, K.} \&
       \bibinfo{author}{Vergassola, M.}
     \newblock \bibinfo{title}{Particles and fields in fluid turbulence}.
     \newblock \emph{\bibinfo{journal}{Rev. Mod. Phys. }}
       \textbf{\bibinfo{volume}{73}}, \bibinfo{pages}{913--975}
       (\bibinfo{year}{2001}).
     
     \bibitem{Bourgoin2006}
     \bibinfo{author}{Bourgoin, M.}, \bibinfo{author}{Ouellette, N.~T.},
       \bibinfo{author}{Xu, H.}, \bibinfo{author}{Berg, J.} \&
       \bibinfo{author}{Bodenschatz, E.}
     \newblock \bibinfo{title}{The role of pair dispersion in turbulent flow}.
     \newblock \emph{\bibinfo{journal}{Science}} \textbf{\bibinfo{volume}{311}},
       \bibinfo{pages}{835--838} (\bibinfo{year}{2006}).
     
     \bibitem{Fouxon2003}
     \bibinfo{author}{Fouxon, A.} \& \bibinfo{author}{Lebedev, V.}
     \newblock \bibinfo{title}{Spectra of turbulence in dilute polymer solutions}.
     \newblock \emph{\bibinfo{journal}{Phys. Fluids}}
       \textbf{\bibinfo{volume}{15}}, \bibinfo{pages}{2060} (\bibinfo{year}{2003}).
     
     \bibitem{Berti2008}
     \bibinfo{author}{Berti, S.}, \bibinfo{author}{Bistagnino, A.},
       \bibinfo{author}{Boffetta, G.}, \bibinfo{author}{Celani, A.} \&
       \bibinfo{author}{Musacchio, S.}
     \newblock \bibinfo{title}{Two-dimensional elastic turbulence}.
     \newblock \emph{\bibinfo{journal}{Phys. Rev. E}} \textbf{\bibinfo{volume}{77}}
       (\bibinfo{year}{2008}).
     
     \bibitem{Shraiman2000}
     \bibinfo{author}{Shraiman, B.~I.} \& \bibinfo{author}{Siggia, E.~D.}
     \newblock \emph{\bibinfo{journal}{Nature}} \textbf{\bibinfo{volume}{405}},
       \bibinfo{pages}{639--646} (\bibinfo{year}{2000}).
     
     \bibitem{Afik2015}
     \bibinfo{author}{Afik, E.}
     \newblock \bibinfo{title}{Robust and highly performant ring detection algorithm
       for 3d particle tracking using 2d microscope imaging}.
       \newblock \emph{\bibinfo{journal}{Sci. Rep. }}
       \textbf{\bibinfo{volume}{5}}, \bibinfo{pages}{13584};
       \newblock \bibinfo{doi}{\href{https://www.nature.com/articles/srep13584}{doi: 10.1038/srep13584}} 
       (\bibinfo{year}{2015}).

     \bibitem{Afik2017b}
     \bibinfo{author}{Afik, E.} \& \bibinfo{author}{Steinberg, V.}
     \newblock \bibinfo{title}{A Lagrangian approach to elastic turbulence in a
     curvilinear microfluidic channel}.
     \newblock \emph{\bibinfo{journal}{figshare}}
     \newblock \bibinfo{doi}{\href{http://dx.doi.org/10.6084/m9.figshare.5112991}{doi:
     10.6084/m9.figshare.5112991}} (\bibinfo{year}{2017}).
     
     \bibitem{Jun2011}
     \bibinfo{author}{Jun, Y.} \& \bibinfo{author}{Steinberg, V.}
     \newblock \bibinfo{title}{Elastic turbulence in a curvilinear channel flow}.
     \newblock \emph{\bibinfo{journal}{Phys. Rev. E}} \textbf{\bibinfo{volume}{84}}
       (\bibinfo{year}{2011}).
     
     \bibitem{Paladin1987}
     \bibinfo{author}{Paladin, G.} \& \bibinfo{author}{Vulpiani, A.}
     \newblock \bibinfo{title}{Anomalous scaling laws in multifractal objects}.
     \newblock \emph{\bibinfo{journal}{Phys. Rep. }}
       \textbf{\bibinfo{volume}{156}}, \bibinfo{pages}{147--225}
       (\bibinfo{year}{1987}).
     
     \bibitem{Cencini2010}
     \bibinfo{author}{Cencini, M.}, \bibinfo{author}{Cecconi, F.} \&
       \bibinfo{author}{Vulpiani, A.}
     \newblock \emph{\bibinfo{title}{Chaos: From Simple Models to Complex Systems}}.
     \newblock Series on advances in statistical mechanics
       (\bibinfo{publisher}{World Scientific}, \bibinfo{year}{2010}).
     
     \bibitem{Frishman2015}
     \bibinfo{author}{Frishman, A.}, \bibinfo{author}{Boffetta, G.},
       \bibinfo{author}{De~Lillo, F.} \& \bibinfo{author}{Liberzon, A.}
     \newblock \bibinfo{title}{Statistical conservation law in two- and
       three-dimensional turbulent flows}.
     \newblock \emph{\bibinfo{journal}{Phys. Rev. E}} \textbf{\bibinfo{volume}{91}}
       (\bibinfo{year}{2015}).
     
     \bibitem{Bitane2012}
     \bibinfo{author}{Bitane, R.}, \bibinfo{author}{Homann, H.} \&
       \bibinfo{author}{Bec, J.}
     \newblock \bibinfo{title}{Time scales of turbulent relative dispersion}.
     \newblock \emph{\bibinfo{journal}{Phys. Rev. E}} \textbf{\bibinfo{volume}{86}}
       (\bibinfo{year}{2012}).
     
     \bibitem{Ouellette2006a}
     \bibinfo{author}{Ouellette, N.~T.}, \bibinfo{author}{Xu, H.},
       \bibinfo{author}{Bourgoin, M.} \& \bibinfo{author}{Bodenschatz, E.}
     \newblock \bibinfo{title}{An experimental study of turbulent relative
       dispersion models}.
     \newblock \emph{\bibinfo{journal}{New J. Phys. }}
       \textbf{\bibinfo{volume}{8}}, \bibinfo{pages}{109} (\bibinfo{year}{2006}).
     
     \bibitem{Yeung2004}
     \bibinfo{author}{Yeung, P.~K.} \& \bibinfo{author}{Borgas, M.~S.}
     \newblock \bibinfo{title}{Relative dispersion in isotropic turbulence. part 1.
       direct numerical simulations and reynolds-number dependence}.
     \newblock \emph{\bibinfo{journal}{J. Fluid Mech. }}
       \textbf{\bibinfo{volume}{503}}, \bibinfo{pages}{93--124}
       (\bibinfo{year}{2004}).
     
     \bibitem{Jullien2003}
     \bibinfo{author}{Jullien, M.-C.}
     \newblock \bibinfo{title}{Dispersion of passive tracers in the direct enstrophy
       cascade: Experimental observations}.
       \newblock \emph{\bibinfo{journal}{Phys. Fluids}}
       \textbf{\bibinfo{volume}{15}}, \bibinfo{pages}{2228} (\bibinfo{year}{2003}).
     
     \bibitem{Ni2013}
     \bibinfo{author}{Ni, R.} \& \bibinfo{author}{Xia, K.-Q.}
     \newblock \bibinfo{title}{Experimental investigation of pair dispersion with
       small initial separation in convective turbulent flows}.
     \newblock \emph{\bibinfo{journal}{Phys. Rev. E}} \textbf{\bibinfo{volume}{87}}
       (\bibinfo{year}{2013}).
     
     \bibitem{Liu2009}
     \bibinfo{author}{Liu, Y.}, \bibinfo{author}{Jun, Y.} \&
       \bibinfo{author}{Steinberg, V.}
     \newblock \bibinfo{title}{Concentration dependence of the longest relaxation
       times of dilute and semi-dilute polymer solutions}.
       \newblock \emph{\bibinfo{journal}{J. Rheol. }}
       \textbf{\bibinfo{volume}{53}}, \bibinfo{pages}{1069--1085}
       (\bibinfo{year}{2009}).
     
     \bibitem{Straw2009}
     \bibinfo{author}{Straw, A.~D.} \& \bibinfo{author}{Dickinson, M.~H.}
     \newblock \bibinfo{title}{Motmot, an open-source toolkit for realtime video
       acquisition and analysis}.
     \newblock \emph{\bibinfo{journal}{Source Code Biol. Med. }}
       \textbf{\bibinfo{volume}{4}}, \bibinfo{pages}{5} (\bibinfo{year}{2009}).
     
     \bibitem{Kelley2011}
     \bibinfo{author}{Kelley, D.~H.} \& \bibinfo{author}{Ouellette, N.~T.}
     \newblock \bibinfo{title}{Using particle tracking to measure flow instabilities
       in an undergraduate laboratory experiment}.
     \newblock \emph{\bibinfo{journal}{Am. J. Phys. }} \textbf{\bibinfo{volume}{79}},
       \bibinfo{pages}{267} (\bibinfo{year}{2011}).
     
     \bibitem{SciPy}
     \bibinfo{author}{Jones, E.}, \bibinfo{author}{Oliphant, T.},
       \bibinfo{author}{Peterson, P.} \emph{et~al.}
     \newblock \bibinfo{title}{\href{http://www.scipy.org/}%
           {{SciPy}: Open source scientific tools for {Python}}}.
       (\bibinfo{year}{2001--}).
     
     \bibitem{Wasserm2007}
     \bibinfo{author}{Wasserman, L.}
     \newblock \emph{\bibinfo{title}{All of Nonparametric Statistics (Springer Texts
       in Statistics)}} (\bibinfo{publisher}{Springer}, \bibinfo{year}{2007}).
     
     \bibitem{Ahnert2007}
     \bibinfo{author}{Ahnert, K.} \& \bibinfo{author}{Abel, M.}
     \newblock \bibinfo{title}{Numerical differentiation of experimental data: local
       versus global methods}.
       \newblock \emph{\bibinfo{journal}{Comput. Phys. Commun. }}
       \textbf{\bibinfo{volume}{177}}, \bibinfo{pages}{764--774}
       (\bibinfo{year}{2007}).
     
     \bibitem{Krakauer2014}
     \bibinfo{author}{Krakauer, N.~Y.} \& \bibinfo{author}{Fekete, B.~M.}
     \newblock \bibinfo{title}{Are climate model simulations useful for forecasting
       precipitation trends? hindcast and synthetic-data experiments}.
       \newblock \emph{\bibinfo{journal}{Environ. Res. Lett. }}
       \textbf{\bibinfo{volume}{9}}, \bibinfo{pages}{024009} (\bibinfo{year}{2014}).
     
     \bibitem{Perez2011}
     \bibinfo{author}{Pérez, F.}, \bibinfo{author}{Granger, B.~E.} \&
       \bibinfo{author}{Hunter, J.~D.}
     \newblock \bibinfo{title}{Python: An ecosystem for scientific computing}.
     \newblock \emph{\bibinfo{journal}{Comput. Sci. Eng. }}
       \textbf{\bibinfo{volume}{13}}, \bibinfo{pages}{13--21}
       (\bibinfo{year}{2011}).
     
     \bibitem{Perez2007}
     \bibinfo{author}{Perez, F.} \& \bibinfo{author}{Granger, B.~E.}
     \newblock \bibinfo{title}{{IP}ython: A system for interactive scientific
       computing}.
     \newblock \emph{\bibinfo{journal}{Comput. Sci. Eng. }}
       \textbf{\bibinfo{volume}{9}}, \bibinfo{pages}{21--29} (\bibinfo{year}{2007}).
     
     \bibitem{McKinney2010}
     \bibinfo{author}{McKinney, W.}
     \newblock \bibinfo{title}{Data structures for statistical computing in python}.
     \newblock In \bibinfo{editor}{van~der Walt, S.} \& \bibinfo{editor}{Millman,
       J.} (eds.) \emph{\bibinfo{booktitle}{Proceedings of the 9th Python in Science
       Conference}}, \bibinfo{pages}{51 -- 56} (\bibinfo{year}{2010}).
     
     \bibitem{Hunter2007}
     \bibinfo{author}{Hunter, J.}
     \newblock \bibinfo{title}{Matplotlib: A 2d graphics environment}.
     \newblock \emph{\bibinfo{journal}{Comput. Sci. Eng. }}
       \textbf{\bibinfo{volume}{9}}, \bibinfo{pages}{90 --95}
       (\bibinfo{year}{2007}).
     
  \end{thebibliography}

\begin{thebibliography}{10}
    \expandafter\ifx\csname url\endcsname\relax
      \def\url#1{\texttt{#1}}\fi
    \expandafter\ifx\csname urlprefix\endcsname\relax\def\urlprefix{URL }\fi
    \providecommand{\bibinfo}[2]{#2}
    \providecommand{\eprint}[2][]{\url{#2}}
    
    \bibitem{_Jun2011}
    \bibinfo{author}{Jun, Y.} \& \bibinfo{author}{Steinberg, V.}
    \newblock \bibinfo{title}{Elastic turbulence in a curvilinear channel flow}.
    \newblock \emph{\bibinfo{journal}{Phys. Rev. E}} \textbf{\bibinfo{volume}{84}}
      (\bibinfo{year}{2011}).
    
    \bibitem{_Falkovich2001}
    \bibinfo{author}{Falkovich, G.}, \bibinfo{author}{Gaw\c{e}dzki, K.} \&
      \bibinfo{author}{Vergassola, M.}
    \newblock \bibinfo{title}{Particles and fields in fluid turbulence}.
    \newblock \emph{\bibinfo{journal}{Rev. Mod. Phys. }}
      \textbf{\bibinfo{volume}{73}}, \bibinfo{pages}{913--975}
      (\bibinfo{year}{2001}).
    
    \bibitem{_Bourgoin2006}
    \bibinfo{author}{Bourgoin, M.}, \bibinfo{author}{Ouellette, N.~T.},
      \bibinfo{author}{Xu, H.}, \bibinfo{author}{Berg, J.} \&
      \bibinfo{author}{Bodenschatz, E.}
    \newblock \bibinfo{title}{The role of pair dispersion in turbulent flow}.
    \newblock \emph{\bibinfo{journal}{Science}} \textbf{\bibinfo{volume}{311}},
      \bibinfo{pages}{835--838} (\bibinfo{year}{2006}).
    
    \bibitem{_Salazar2009}
    \bibinfo{author}{Salazar, J.~P.} \& \bibinfo{author}{Collins, L.~R.}
    \newblock \bibinfo{title}{Two-particle dispersion in isotropic turbulent
      flows}.
    \newblock \emph{\bibinfo{journal}{Annu. Rev. Fluid Mech. }}
      \textbf{\bibinfo{volume}{41}}, \bibinfo{pages}{405--432}
      (\bibinfo{year}{2009}).
    
    \bibitem{_Batchelor1952}
    \bibinfo{author}{Batchelor, G.~K.}
    \newblock \bibinfo{title}{The effect of homogeneous turbulence on material
      lines and surfaces}.
      \newblock \emph{\bibinfo{journal}{Proc. R. Soc. A}}
      \textbf{\bibinfo{volume}{213}}, \bibinfo{pages}{349--366}
      (\bibinfo{year}{1952}).
    
    \bibitem{_Monin2007b}
    \bibinfo{author}{Monin, A.~S.} \& \bibinfo{author}{Yaglom, A.~M.}
    \newblock \emph{\bibinfo{title}{Statistical Fluid Mechanics, Volume II:
      Mechanics of Turbulence (Dover Books on Physics)}}, vol.~\bibinfo{volume}{2}
      (\bibinfo{publisher}{Dover Publications}, \bibinfo{year}{2007}).
    \newblock
    
    \bibitem{_Luthi2005}
    \bibinfo{author}{L\"{u}thi, B.}, \bibinfo{author}{Tsinober, A.} \&
      \bibinfo{author}{Kinzelbach, W.}
    \newblock \bibinfo{title}{Lagrangian measurement of vorticity dynamics in
      turbulent flow}.
    \newblock \emph{\bibinfo{journal}{J. Fluid Mech. }}
      \textbf{\bibinfo{volume}{528}}, \bibinfo{pages}{87--118}
      (\bibinfo{year}{2005}).
    
    \bibitem{_Guala2005}
    \bibinfo{author}{Guala, M.}, \bibinfo{author}{L\"{u}thi, B.},
      \bibinfo{author}{Liberzon, A.}, \bibinfo{author}{Tsinober, A.} \&
      \bibinfo{author}{Kinzelbach, W.}
    \newblock \bibinfo{title}{On the evolution of material lines and vorticity in
      homogeneous turbulence}.
    \newblock \emph{\bibinfo{journal}{J. Fluid Mech. }}
      \textbf{\bibinfo{volume}{533}} (\bibinfo{year}{2005}).
    
    \bibitem{_Cencini2010}
    \bibinfo{author}{Cencini, M.}, \bibinfo{author}{Cecconi, F.} \&
      \bibinfo{author}{Vulpiani, A.}
    \newblock \emph{\bibinfo{title}{Chaos: From Simple Models to Complex Systems}}.
    \newblock Series on advances in statistical mechanics
      (\bibinfo{publisher}{World Scientific}, \bibinfo{year}{2010}).
    
    \bibitem{_Paladin1987}
    \bibinfo{author}{Paladin, G.} \& \bibinfo{author}{Vulpiani, A.}
    \newblock \bibinfo{title}{Anomalous scaling laws in multifractal objects}.
    \newblock \emph{\bibinfo{journal}{Phys. Rep. }}
      \textbf{\bibinfo{volume}{156}}, \bibinfo{pages}{147--225}
      (\bibinfo{year}{1987}).
    
    \bibitem{_Fouxon2003}
    \bibinfo{author}{Fouxon, A.} \& \bibinfo{author}{Lebedev, V.}
    \newblock \bibinfo{title}{Spectra of turbulence in dilute polymer solutions}.
    \newblock \emph{\bibinfo{journal}{Phys. Fluids}}
      \textbf{\bibinfo{volume}{15}}, \bibinfo{pages}{2060} (\bibinfo{year}{2003}).
    
    \bibitem{_Berti2008}
    \bibinfo{author}{Berti, S.}, \bibinfo{author}{Bistagnino, A.},
      \bibinfo{author}{Boffetta, G.}, \bibinfo{author}{Celani, A.} \&
      \bibinfo{author}{Musacchio, S.}
    \newblock \bibinfo{title}{Two-dimensional elastic turbulence}.
    \newblock \emph{\bibinfo{journal}{Phys. Rev. E}} \textbf{\bibinfo{volume}{77}}
      (\bibinfo{year}{2008}).
    
    \bibitem{_Steinberg2009}
    \bibinfo{author}{Steinberg, V.}
    \newblock \bibinfo{title}{Elastic stresses in random flow of a dilute polymer
      solution and the turbulent drag reduction problem}.
    \newblock \emph{\bibinfo{journal}{C R Phys}}
      \textbf{\bibinfo{volume}{10}}, \bibinfo{pages}{728--739}
      (\bibinfo{year}{2009}).
    
    \bibitem{_Burghelea2004a}
    \bibinfo{author}{Burghelea, T.}, \bibinfo{author}{Segre, E.} \&
      \bibinfo{author}{Steinberg, V.}
    \newblock \bibinfo{title}{Statistics of particle pair separations in the
      elastic turbulent flow of a dilute polymer solution}.
      \newblock \emph{\bibinfo{journal}{Europhys. Lett.}}
      \textbf{\bibinfo{volume}{68}}, \bibinfo{pages}{529--535}
      (\bibinfo{year}{2004}).
    
    \bibitem{_Burghelea2007}
    \bibinfo{author}{Burghelea, T.}, \bibinfo{author}{Segre, E.} \&
      \bibinfo{author}{Steinberg, V.}
    \newblock \bibinfo{title}{Elastic turbulence in von karman swirling flow
      between two disks}.
    \newblock \emph{\bibinfo{journal}{Phys. Fluids}}
      \textbf{\bibinfo{volume}{19}}, \bibinfo{pages}{053104}
      (\bibinfo{year}{2007}).
    
    \bibitem{_Gerashchenko2005}
    \bibinfo{author}{Gerashchenko, S.}, \bibinfo{author}{Chevallard, C.} \&
      \bibinfo{author}{Steinberg, V.}
    \newblock \bibinfo{title}{Single polymer dynamics: coil-stretch transition in a
      random flow}.
    \newblock \emph{\bibinfo{journal}{Europhys. Lett.}}
      \textbf{\bibinfo{volume}{71}}, \bibinfo{pages}{221--227}
      (\bibinfo{year}{2005}).
    
    \bibitem{_Jullien2003}
    \bibinfo{author}{Jullien, M.-C.}
    \newblock \bibinfo{title}{Dispersion of passive tracers in the direct enstrophy
      cascade: Experimental observations}.
    \newblock \emph{\bibinfo{journal}{Phys. Fluids}}
      \textbf{\bibinfo{volume}{15}}, \bibinfo{pages}{2228} (\bibinfo{year}{2003}).
    
    \bibitem{_Smith1999}
    \bibinfo{author}{Smith, D.~E.}
    \newblock \bibinfo{title}{Single-polymer dynamics in steady shear flow}.
    \newblock \emph{\bibinfo{journal}{Science}} \textbf{\bibinfo{volume}{283}},
      \bibinfo{pages}{1724--1727} (\bibinfo{year}{1999}).
    
    \bibitem{_Lin1972}
    \bibinfo{author}{Lin, J.-T.}
    \newblock \bibinfo{title}{Relative dispersion in the enstrophy-cascading
      inertial range of homogeneous two-dimensional turbulence}.
    \newblock \emph{\bibinfo{journal}{J. Atmos. Sci.}}
      \textbf{\bibinfo{volume}{29}}, \bibinfo{pages}{394--396}
      (\bibinfo{year}{1972}).
    
    \bibitem{_Ni2013}
    \bibinfo{author}{Ni, R.} \& \bibinfo{author}{Xia, K.-Q.}
    \newblock \bibinfo{title}{Experimental investigation of pair dispersion with
      small initial separation in convective turbulent flows}.
    \newblock \emph{\bibinfo{journal}{Phys. Rev. E}} \textbf{\bibinfo{volume}{87}}
      (\bibinfo{year}{2013}).
    
    \bibitem{_Bewley2013}
    \bibinfo{author}{Bewley, G.~P.}, \bibinfo{author}{Saw, E.-W.} \&
      \bibinfo{author}{Bodenschatz, E.}
    \newblock \bibinfo{title}{Observation of the sling effect}.
    \newblock \emph{\bibinfo{journal}{New J. Phys. }}
      \textbf{\bibinfo{volume}{15}}, \bibinfo{pages}{083051}
      (\bibinfo{year}{2013}).
    
    \bibitem{_Ouellette2006a}
    \bibinfo{author}{Ouellette, N.~T.}, \bibinfo{author}{Xu, H.},
      \bibinfo{author}{Bourgoin, M.} \& \bibinfo{author}{Bodenschatz, E.}
    \newblock \bibinfo{title}{An experimental study of turbulent relative
      dispersion models}.
    \newblock \emph{\bibinfo{journal}{New J. Phys. }}
      \textbf{\bibinfo{volume}{8}}, \bibinfo{pages}{109} (\bibinfo{year}{2006}).
    
  \end{thebibliography}
\end{document}